%
%
%
%

\catcode `\@=11 

\def\@version{1.4}
\def\@verdate{22nd Feb 1994}

%
%
%
%


\newif\ifprod@font

\ifx\@typeface\undefined
  \def\@typeface{Comp. Modern}\prod@fontfalse
\else
  \prod@fonttrue 
\fi

\def\newfam{\alloc@8\fam\chardef\sixt@@n} 

\ifprod@font
\font\fiverm=mtr10 at 5pt
\font\fivebf=mtbx10 at 5pt
\font\fiveit=mtti10 at 5pt
\font\fivesl=mtsl10 at 5pt
\font\fivett=mttt10 at 5pt     \hyphenchar\fivett=-1
\font\fivecsc=mtcsc10 at 5pt
\font\fivesf=mtss10 at 5pt
\font\fivei=mtmi10 at 5pt      \skewchar\fivei='177
\font\fivemib=mtmib10 at 5pt   \skewchar\fivemib='177
\font\fivesy=mtsy10 at 5pt     \skewchar\fivesy='60
\font\fivesyb=mtbsy10 at 5pt   \skewchar\fivesyb='60

\font\sixrm=mtr10 at 6pt
\font\sixbf=mtbx10 at 6pt
\font\sixit=mtti10 at 6pt
\font\sixsl=mtsl10 at 6pt
\font\sixtt=mttt10 at 6pt      \hyphenchar\sixtt=-1
\font\sixcsc=mtcsc10 at 6pt
\font\sixsf=mtss10 at 6pt
\font\sixi=mtmi10 at 6pt       \skewchar\sixi='177
\font\sixmib=mtmib10 at 6pt    \skewchar\sixmib='177
\font\sixsy=mtsy10 at 6pt      \skewchar\sixsy='60
\font\sixsyb=mtbsy10 at 6pt    \skewchar\sixsyb='60

\font\sevenrm=mtr10 at 7pt
\font\sevenbf=mtbx10 at 7pt
\font\sevenit=mtti10 at 7pt
\font\sevensl=mtsl10 at 7pt
\font\seventt=mttt10 at 7pt     \hyphenchar\seventt=-1
\font\sevencsc=mtcsc10 at 7pt
\font\sevensf=mtss10 at 7pt
\font\seveni=mtmi10 at 7pt      \skewchar\seveni='177
\font\sevenmib=mtmib10 at 7pt   \skewchar\sevenmib='177
\font\sevensy=mtsy10 at 7pt     \skewchar\sevensy='60
\font\sevensyb=mtbsy10 at 7pt   \skewchar\sevensyb='60

\font\eightrm=mtr10 at 8pt
\font\eightbf=mtbx10 at 8pt
\font\eightit=mtti10 at 8pt
\font\eighti=mtmi10 at 8pt      \skewchar\eighti='177
\font\eightmib=mtmib10 at 8pt   \skewchar\eightmib='177
\font\eightsy=mtsy10 at 8pt     \skewchar\eightsy='60
\font\eightsyb=mtbsy10 at 8pt   \skewchar\eightsyb='60
\font\eightsl=mtsl10 at 8pt
\font\eighttt=mttt10 at 8pt     \hyphenchar\eighttt=-1
\font\eightcsc=mtcsc10 at 8pt
\font\eightsf=mtss10 at 8pt

\font\ninerm=mtr10 at 9pt
\font\ninebf=mtbx10 at 9pt
\font\nineit=mtti10 at 9pt
\font\ninei=mtmi10 at 9pt      \skewchar\ninei='177
\font\ninemib=mtmib10 at 9pt   \skewchar\ninemib='177
\font\ninesy=mtsy10 at 9pt     \skewchar\ninesy='60
\font\ninesyb=mtbsy10 at 9pt   \skewchar\ninesyb='60
\font\ninesl=mtsl10 at 9pt
\font\ninett=mttt10 at 9pt     \hyphenchar\ninett=-1
\font\ninecsc=mtcsc10 at 9pt
\font\ninesf=mtss10 at 9pt

\font\tenrm=mtr10
\font\tenbf=mtbx10
\font\tenit=mtti10
\font\teni=mtmi10		\skewchar\teni='177
\font\tenmib=mtmib10	\skewchar\tenmib='177
\font\tensy=mtsy10		\skewchar\tensy='60
\font\tensyb=mtbsy10	\skewchar\tensyb='60
\font\tenex=cmex10
\font\tensl=mtsl10
\font\tentt=mttt10		\hyphenchar\tentt=-1
\font\tencsc=mtcsc10
\font\tensf=mtss10

\font\elevenrm=mtr10 at 11pt
\font\elevenbf=mtbx10 at 11pt
\font\elevenit=mtti10 at 11pt
\font\eleveni=mtmi10 at 11pt      \skewchar\eleveni='177
\font\elevenmib=mtmib10 at 11pt   \skewchar\elevenmib='177
\font\elevensy=mtsy10 at 11pt     \skewchar\elevensy='60
\font\elevensyb=mtbsy10 at 11pt   \skewchar\elevensyb='60
\font\elevensl=mtsl10 at 11pt
\font\eleventt=mttt10 at 11pt     \hyphenchar\eleventt=-1
\font\elevencsc=mtcsc10 at 11pt
\font\elevensf=mtss10 at 11pt

\font\twelverm=mtr10 at 12pt
\font\twelvebf=mtbx10 at 12pt
\font\twelveit=mtti10 at 12pt
\font\twelvesl=mtsl10 at 12pt
\font\twelvett=mttt10 at 12pt     \hyphenchar\twelvett=-1
\font\twelvecsc=mtcsc10 at 12pt
\font\twelvesf=mtss10 at 12pt
\font\twelvei=mtmi10 at 12pt      \skewchar\twelvei='177
\font\twelvemib=mtmib10 at 12pt   \skewchar\twelvemib='177
\font\twelvesy=mtsy10 at 12pt     \skewchar\twelvesy='60
\font\twelvesyb=mtbsy10 at 12pt   \skewchar\twelvesyb='60

\font\fourteenrm=mtr10 at 14pt
\font\fourteenbf=mtbx10 at 14pt
\font\fourteenit=mtti10 at 14pt
\font\fourteeni=mtmi10 at 14pt      \skewchar\fourteeni='177
\font\fourteenmib=mtmib10 at 14pt   \skewchar\fourteenmib='177
\font\fourteensy=mtsy10 at 14pt     \skewchar\fourteensy='60
\font\fourteensyb=mtbsy10 at 14pt   \skewchar\fourteensyb='60
\font\fourteensl=mtsl10 at 14pt
\font\fourteentt=mttt10 at 14pt     \hyphenchar\fourteentt=-1
\font\fourteencsc=mtcsc10 at 14pt
\font\fourteensf=mtss10 at 14pt

\font\seventeenrm=mtr10 at 17pt
\font\seventeenbf=mtbx10 at 17pt
\font\seventeenit=mtti10 at 17pt
\font\seventeeni=mtmi10 at 17pt      \skewchar\seventeeni='177
\font\seventeenmib=mtmib10 at 17pt   \skewchar\seventeenmib='177
\font\seventeensy=mtsy10 at 17pt     \skewchar\seventeensy='60
\font\seventeensyb=mtbsy10 at 17pt   \skewchar\seventeensyb='60
\font\seventeensl=mtsl10 at 17pt
\font\seventeentt=mttt10 at 17pt     \hyphenchar\seventeentt=-1
\font\seventeencsc=mtcsc10 at 17pt
\font\seventeensf=mtss10 at 17pt


\newfam\xmfam
\newfam\ymfam

\font\fivexm=mtxm10 at 5pt
\font\sixxm=mtxm10 at 6pt
\font\sevenxm=mtxm10 at 7pt
\font\eightxm=mtxm10 at 8pt
\font\ninexm=mtxm10 at 9pt
\font\tenxm=mtxm10
\font\elevenxm=mtxm10 at 11pt
\font\twelvexm=mtxm10 at 12pt
\font\fourteenxm=mtxm10 at 14pt
\font\seventeenxm=mtxm10 at 17pt

\font\fiveym=mtym10 at 5pt
\font\sixym=mtym10 at 6pt
\font\sevenym=mtym10 at 7pt
\font\eightym=mtym10 at 8pt
\font\nineym=mtym10 at 9pt
\font\tenym=mtym10
\font\elevenym=mtym10 at 11pt
\font\twelveym=mtym10 at 12pt
\font\fourteenym=mtym10 at 14pt
\font\seventeenym=mtym10 at 17pt
\else
\font\fiverm=cmr5
\font\fivei=cmmi5             \skewchar\fivei='177
\font\fivemib=cmmib10 at 5pt  \skewchar\fivemib='177
\font\fivesy=cmsy5            \skewchar\fivesy='60
\font\fivesyb=cmbsy10 at 5pt  \skewchar\fivesyb='60
\font\fivebf=cmbx5

\font\sixrm=cmr6
\font\sixi=cmmi6             \skewchar\sixi='177
\font\sixmib=cmmib10 at 6pt  \skewchar\sixmib='177
\font\sixsy=cmsy6            \skewchar\sixsy='60
\font\sixsyb=cmbsy10 at 6pt  \skewchar\sixsyb='60
\font\sixbf=cmbx6

\font\sevenrm=cmr7
\font\seveni=cmmi7             \skewchar\seveni='177
\font\sevenmib=cmmib10 at 7pt  \skewchar\sevenmib='177
\font\sevensy=cmsy7            \skewchar\sevensy='60
\font\sevensyb=cmbsy10 at 7pt  \skewchar\sevensyb='60
\font\sevenbf=cmbx7

\font\eightrm=cmr8
\font\eightbf=cmbx8
\font\eightit=cmti8
\font\eighti=cmmi8			\skewchar\eighti='177
\font\eightmib=cmmib10 at 8pt	\skewchar\eightmib='177
\font\eightsy=cmsy8			\skewchar\eightsy='60
\font\eightsyb=cmbsy10 at 8pt	\skewchar\eightsyb='60
\font\eightsl=cmsl8
\font\eighttt=cmtt8			\hyphenchar\eighttt=-1
\font\eightcsc=cmcsc10 at 8pt
\font\eightsf=cmss8

\font\ninerm=cmr9
\font\ninebf=cmbx9
\font\nineit=cmti9
\font\ninei=cmmi9			\skewchar\ninei='177
\font\ninemib=cmmib10 at 9pt	\skewchar\ninemib='177
\font\ninesy=cmsy9			\skewchar\ninesy='60
\font\ninesyb=cmbsy10 at 9pt	\skewchar\ninesyb='60
\font\ninesl=cmsl9
\font\ninett=cmtt9			\hyphenchar\ninett=-1
\font\ninecsc=cmcsc10 at 9pt
\font\ninesf=cmss9

\font\tenrm=cmr10
\font\tenbf=cmbx10
\font\tenit=cmti10
\font\teni=cmmi10		\skewchar\teni='177
\font\tenmib=cmmib10	\skewchar\tenmib='177
\font\tensy=cmsy10		\skewchar\tensy='60
\font\tensyb=cmbsy10	\skewchar\tensyb='60
\font\tenex=cmex10
\font\tensl=cmsl10
\font\tentt=cmtt10		\hyphenchar\tentt=-1
\font\tencsc=cmcsc10
\font\tensf=cmss10

\font\elevenrm=cmr10 scaled \magstephalf
\font\elevenbf=cmbx10 scaled \magstephalf
\font\elevenit=cmti10 scaled \magstephalf
\font\eleveni=cmmi10 scaled \magstephalf	\skewchar\eleveni='177
\font\elevenmib=cmmib10 scaled \magstephalf	\skewchar\elevenmib='177
\font\elevensy=cmsy10 scaled \magstephalf	\skewchar\elevensy='60
\font\elevensyb=cmbsy10 scaled \magstephalf	\skewchar\elevensyb='60
\font\elevensl=cmsl10 scaled \magstephalf
\font\eleventt=cmtt10 scaled \magstephalf	\hyphenchar\eleventt=-1
\font\elevencsc=cmcsc10 scaled \magstephalf
\font\elevensf=cmss10 scaled \magstephalf

\font\twelverm=cmr10 scaled \magstep1
\font\twelvebf=cmbx10 scaled \magstep1
\font\twelvei=cmmi10 scaled \magstep1      \skewchar\twelvei='177
\font\twelvemib=cmmib10 scaled \magstep1   \skewchar\twelvemib='177
\font\twelvesy=cmsy10 scaled \magstep1     \skewchar\twelvesy='60
\font\twelvesyb=cmbsy10 scaled \magstep1   \skewchar\twelvesyb='60

\font\fourteenrm=cmr10 scaled \magstep2
\font\fourteenbf=cmbx10 scaled \magstep2
\font\fourteenit=cmti10 scaled \magstep2
\font\fourteeni=cmmi10 scaled \magstep2		\skewchar\fourteeni='177
\font\fourteenmib=cmmib10 scaled \magstep2	\skewchar\fourteenmib='177
\font\fourteensy=cmsy10 scaled \magstep2	\skewchar\fourteensy='60
\font\fourteensyb=cmbsy10 scaled \magstep2	\skewchar\fourteensyb='60
\font\fourteensl=cmsl10 scaled \magstep2
\font\fourteentt=cmtt10 scaled \magstep2	\hyphenchar\fourteentt=-1
\font\fourteencsc=cmcsc10 scaled \magstep2
\font\fourteensf=cmss10 scaled \magstep2

\font\seventeenrm=cmr10 scaled \magstep3
\font\seventeenbf=cmbx10 scaled \magstep3
\font\seventeenit=cmti10 scaled \magstep3
\font\seventeeni=cmmi10 scaled \magstep3	\skewchar\seventeeni='177
\font\seventeenmib=cmmib10 scaled \magstep3	\skewchar\seventeenmib='177
\font\seventeensy=cmsy10 scaled \magstep3	\skewchar\seventeensy='60
\font\seventeensyb=cmbsy10 scaled \magstep3	\skewchar\seventeensyb='60
\font\seventeensl=cmsl10 scaled \magstep3
\font\seventeentt=cmtt10 scaled \magstep3	\hyphenchar\seventeentt=-1
\font\seventeencsc=cmcsc10 scaled \magstep3
\font\seventeensf=cmss10 scaled \magstep3
\fi

\def\hexnumber#1{\ifcase#1 0\or1\or2\or3\or4\or5\or6\or7\or8\or9\or
  A\or B\or C\or D\or E\or F\fi}

\ifprod@font
  \edef\@xm{\hexnumber\xmfam}
  \edef\@ym{\hexnumber\ymfam}
\fi

\def\makestrut{%
  \setbox\strutbox=\hbox{%
    \vrule height.7\baselineskip depth.3\baselineskip width \z@}%
}

\def\baselinestretch{1}
\newskip\tmp@bls

\def\b@ls#1{
  \tmp@bls=#1\relax
  \baselineskip=#1\relax\makestrut
  \normalbaselineskip=\baselinestretch\tmp@bls
  \normalbaselines
}

\def\nostb@ls#1{
  \normalbaselineskip=#1\relax
  \normalbaselines
  \makestrut
}

%

\newfam\mibfam 
\newfam\sybfam 
\newfam\scfam  
\newfam\sffam  

\def\mit{\fam\@ne}

\def\cal{\fam\tw@}

\def\em{\ifdim\fontdimen1\font>\z@ \rm\else\it\fi}

\textfont3=\tenex
\scriptfont3=\tenex
\scriptscriptfont3=\tenex

\setbox0=\hbox{\tenex B} \p@renwd=\wd0 

\def\eightpoint{
  \def\rm{\fam0\eightrm}%
  \textfont0=\eightrm \scriptfont0=\sixrm \scriptscriptfont0=\fiverm%
  \textfont1=\eighti  \scriptfont1=\sixi  \scriptscriptfont1=\fivei%
  \textfont2=\eightsy \scriptfont2=\sixsy \scriptscriptfont2=\fivesy%
  \textfont\itfam=\eightit\def\it{\fam\itfam\eightit}%
  \ifprod@font
    \scriptfont\itfam=\sixit
      \scriptscriptfont\itfam=\fiveit
  \else
    \scriptfont\itfam=\eightit
      \scriptscriptfont\itfam=\eightit
  \fi
  \textfont\bffam=\eightbf%
    \scriptfont\bffam=\sixbf%
      \scriptscriptfont\bffam=\fivebf%
  \def\bf{\fam\bffam\eightbf}%
  \textfont\slfam=\eightsl\def\sl{\fam\slfam\eightsl}%
  \ifprod@font
    \scriptfont\slfam=\sixsl
      \scriptscriptfont\slfam=\fivesl
  \else
    \scriptfont\slfam=\eightsl
      \scriptscriptfont\slfam=\eightsl
  \fi
  \textfont\ttfam=\eighttt\def\tt{\fam\ttfam\eighttt}%
  \ifprod@font
    \scriptfont\ttfam=\sixtt
      \scriptscriptfont\ttfam=\fivett
  \else
    \scriptfont\ttfam=\eighttt
      \scriptscriptfont\ttfam=\eighttt
  \fi
  \textfont\scfam=\eightcsc\def\sc{\fam\scfam\eightcsc}%
  \ifprod@font
    \scriptfont\scfam=\sixcsc
      \scriptscriptfont\scfam=\fivecsc
  \else
    \scriptfont\scfam=\eightcsc
      \scriptscriptfont\scfam=\eightcsc
  \fi
  \textfont\sffam=\eightsf\def\sf{\fam\sffam\eightsf}%
  \ifprod@font
    \scriptfont\sffam=\sixsf
      \scriptscriptfont\sffam=\fivesf
  \else
    \scriptfont\sffam=\eightsf
      \scriptscriptfont\sffam=\eightsf
  \fi
  \textfont\mibfam=\eightmib
    \scriptfont\mibfam=\sixmib
      \scriptscriptfont\mibfam=\fivemib
  \textfont\sybfam=\eightsyb
    \scriptfont\sybfam=\sixsyb
      \scriptscriptfont\sybfam=\fivesyb
  \ifprod@font
    \textfont\xmfam=\eightxm
      \scriptfont\xmfam=\sixxm
        \scriptscriptfont\xmfam=\fivexm
    \textfont\ymfam=\eightym
      \scriptfont\ymfam=\sixym
        \scriptscriptfont\ymfam=\fiveym
  \fi
  \def\oldstyle{\fam\@ne\eighti}%
  \def\boldstyle{\fam\mibfam\eightmib}%
  \b@ls{10pt}\rm%
}

\def\ninepoint{
  \def\rm{\fam0\ninerm}%
  \textfont0=\ninerm \scriptfont0=\sixrm \scriptscriptfont0=\fiverm%
  \textfont1=\ninei  \scriptfont1=\sixi  \scriptscriptfont1=\fivei%
  \textfont2=\ninesy \scriptfont2=\sixsy \scriptscriptfont2=\fivesy%
  \textfont\itfam=\nineit\def\it{\fam\itfam\nineit}%
  \ifprod@font
    \scriptfont\itfam=\sixit
      \scriptscriptfont\itfam=\fiveit
  \else
    \scriptfont\itfam=\nineit
      \scriptscriptfont\itfam=\nineit
  \fi
  \textfont\bffam=\ninebf%
    \scriptfont\bffam=\sixbf%
      \scriptscriptfont\bffam=\fivebf%
  \def\bf{\fam\bffam\ninebf}%
  \textfont\slfam=\ninesl\def\sl{\fam\slfam\ninesl}%
  \ifprod@font
    \scriptfont\slfam=\sixsl
      \scriptscriptfont\slfam=\fivesl
  \else
    \scriptfont\slfam=\ninesl
      \scriptscriptfont\slfam=\ninesl
  \fi
  \textfont\ttfam=\ninett\def\tt{\fam\ttfam\ninett}%
  \ifprod@font
    \scriptfont\ttfam=\sixtt
      \scriptscriptfont\ttfam=\fivett
  \else
    \scriptfont\ttfam=\ninett
      \scriptscriptfont\ttfam=\ninett
  \fi
  \textfont\scfam=\ninecsc\def\sc{\fam\scfam\ninecsc}%
  \ifprod@font
    \scriptfont\scfam=\sixcsc
      \scriptscriptfont\scfam=\fivecsc
  \else
    \scriptfont\scfam=\ninecsc
      \scriptscriptfont\scfam=\ninecsc
  \fi
  \textfont\sffam=\ninesf\def\sf{\fam\sffam\ninesf}%
  \ifprod@font
    \scriptfont\sffam=\sixsf
      \scriptscriptfont\sffam=\fivesf
  \else
    \scriptfont\sffam=\ninesf
      \scriptscriptfont\sffam=\ninesf
  \fi
  \textfont\mibfam=\ninemib
    \scriptfont\mibfam=\sixmib
      \scriptscriptfont\mibfam=\fivemib
  \textfont\sybfam=\ninesyb
    \scriptfont\sybfam=\sixsyb
      \scriptscriptfont\sybfam=\fivesyb
  \ifprod@font
    \textfont\xmfam=\ninexm
      \scriptfont\xmfam=\sixxm
        \scriptscriptfont\xmfam=\fivexm
    \textfont\ymfam=\nineym
      \scriptfont\ymfam=\sixym
        \scriptscriptfont\ymfam=\fiveym
  \fi
  \def\oldstyle{\fam\@ne\ninei}%
  \def\boldstyle{\fam\mibfam\ninemib}%
  \b@ls{\TextLeading plus \Feathering}\rm%
}

\def\tenpoint{
  \def\rm{\fam0\tenrm}%
  \textfont0=\tenrm \scriptfont0=\sevenrm \scriptscriptfont0=\fiverm%
  \textfont1=\teni  \scriptfont1=\seveni  \scriptscriptfont1=\fivei%
  \textfont2=\tensy \scriptfont2=\sevensy \scriptscriptfont2=\fivesy%
  \textfont\itfam=\tenit\def\it{\fam\itfam\tenit}%
  \ifprod@font
    \scriptfont\itfam=\sevenit
      \scriptscriptfont\itfam=\fiveit
  \else
    \scriptfont\itfam=\tenit
      \scriptscriptfont\itfam=\tenit
  \fi
  \textfont\bffam=\tenbf%
    \scriptfont\bffam=\sevenbf%
      \scriptscriptfont\bffam=\fivebf%
  \def\bf{\fam\bffam\tenbf}%
  \textfont\slfam=\tensl\def\sl{\fam\slfam\tensl}%
  \ifprod@font
    \scriptfont\slfam=\sevensl
      \scriptscriptfont\slfam=\fivesl
  \else
    \scriptfont\slfam=\tensl
      \scriptscriptfont\slfam=\tensl
  \fi
  \textfont\ttfam=\tentt\def\tt{\fam\ttfam\tentt}%
  \ifprod@font
    \scriptfont\ttfam=\seventt
      \scriptscriptfont\ttfam=\fivett
  \else
    \scriptfont\ttfam=\tentt
      \scriptscriptfont\ttfam=\tentt
  \fi
  \textfont\scfam=\tencsc\def\sc{\fam\scfam\tencsc}%
  \ifprod@font
    \scriptfont\scfam=\sevencsc
      \scriptscriptfont\scfam=\fivecsc
  \else
    \scriptfont\scfam=\tencsc
      \scriptscriptfont\scfam=\tencsc
  \fi
  \textfont\sffam=\tensf\def\sf{\fam\sffam\tensf}%
  \ifprod@font
    \scriptfont\sffam=\sevensf
      \scriptscriptfont\sffam=\fivesf
  \else
    \scriptfont\sffam=\tensf
      \scriptscriptfont\sffam=\tensf
  \fi
  \textfont\mibfam=\tenmib
    \scriptfont\mibfam=\sevenmib
      \scriptscriptfont\mibfam=\fivemib
  \textfont\sybfam=\tensyb
    \scriptfont\sybfam=\sevensyb
      \scriptscriptfont\sybfam=\fivesyb
  \ifprod@font
    \textfont\xmfam=\tenxm
      \scriptfont\xmfam=\sevenxm
        \scriptscriptfont\xmfam=\fivexm
    \textfont\ymfam=\tenym
      \scriptfont\ymfam=\sevenym
        \scriptscriptfont\ymfam=\fiveym
  \fi
  \def\oldstyle{\fam\@ne\teni}%
  \def\boldstyle{\fam\mibfam\tenmib}%
  \b@ls{11pt}\rm%
}

\def\elevenpoint{
  \def\rm{\fam0\elevenrm}%
  \textfont0=\elevenrm \scriptfont0=\eightrm \scriptscriptfont0=\sixrm%
  \textfont1=\eleveni  \scriptfont1=\eighti  \scriptscriptfont1=\sixi%
  \textfont2=\elevensy \scriptfont2=\eightsy \scriptscriptfont2=\sixsy%
  \textfont\itfam=\elevenit\def\it{\fam\itfam\elevenit}%
  \ifprod@font
    \scriptfont\itfam=\eightit
      \scriptscriptfont\itfam=\sixit
  \else
    \scriptfont\itfam=\elevenit
      \scriptscriptfont\itfam=\elevenit
  \fi
  \textfont\bffam=\elevenbf%
    \scriptfont\bffam=\eightbf%
      \scriptscriptfont\bffam=\sixbf%
  \def\bf{\fam\bffam\elevenbf}%
  \textfont\slfam=\elevensl\def\sl{\fam\slfam\elevensl}%
  \ifprod@font
    \scriptfont\slfam=\eightsl
      \scriptscriptfont\slfam=\sixsl
  \else
    \scriptfont\slfam=\elevensl
      \scriptscriptfont\slfam=\elevensl
  \fi
  \textfont\ttfam=\eleventt\def\tt{\fam\ttfam\eleventt}%
  \ifprod@font
    \scriptfont\ttfam=\eighttt
      \scriptscriptfont\ttfam=\sixtt
  \else
    \scriptfont\ttfam=\eleventt
      \scriptscriptfont\ttfam=\eleventt
  \fi
  \textfont\scfam=\elevencsc\def\sc{\fam\scfam\elevencsc}%
  \ifprod@font
    \scriptfont\scfam=\eightcsc
      \scriptscriptfont\scfam=\sixcsc
  \else
    \scriptfont\scfam=\elevencsc
      \scriptscriptfont\scfam=\elevencsc
  \fi
  \textfont\sffam=\elevensf\def\sf{\fam\sffam\elevensf}%
  \ifprod@font
    \scriptfont\sffam=\eightsf
      \scriptscriptfont\sffam=\sixsf
  \else
    \scriptfont\sffam=\elevensf
      \scriptscriptfont\sffam=\elevensf
  \fi
  \textfont\mibfam=\elevenmib
    \scriptfont\mibfam=\eightmib
      \scriptscriptfont\mibfam=\sixmib
  \textfont\sybfam=\elevensyb
    \scriptfont\sybfam=\eightsyb
      \scriptscriptfont\sybfam=\sixsyb
  \ifprod@font
    \textfont\xmfam=\elevenxm
      \scriptfont\xmfam=\eightxm
       \scriptscriptfont\xmfam=\sixxm
    \textfont\ymfam=\elevenym
      \scriptfont\ymfam=\eightym
        \scriptscriptfont\ymfam=\sixym
   \fi
  \def\oldstyle{\fam\@ne\eleveni}%
  \def\boldstyle{\fam\mibfam\elevenmib}%
  \b@ls{13pt}\rm%
}

\def\fourteenpoint{
  \def\rm{\fam0\fourteenrm}%
  \textfont0\fourteenrm  \scriptfont0\tenrm  \scriptscriptfont0\sevenrm%
  \textfont1\fourteeni   \scriptfont1\teni   \scriptscriptfont1\seveni%
  \textfont2\fourteensy  \scriptfont2\tensy  \scriptscriptfont2\sevensy%
  \textfont\itfam=\fourteenit\def\it{\fam\itfam\fourteenit}%
  \ifprod@font
    \scriptfont\itfam=\tenit
      \scriptscriptfont\itfam=\sevenit
  \else
    \scriptfont\itfam=\fourteenit
      \scriptscriptfont\itfam=\fourteenit
  \fi
  \textfont\bffam=\fourteenbf%
    \scriptfont\bffam=\tenbf%
      \scriptscriptfont\bffam=\sevenbf%
  \def\bf{\fam\bffam\fourteenbf}%
  \textfont\slfam=\fourteensl\def\sl{\fam\slfam\fourteensl}%
  \ifprod@font
    \scriptfont\slfam=\tensl
      \scriptscriptfont\slfam=\sevensl
  \else
    \scriptfont\slfam=\fourteensl
      \scriptscriptfont\slfam=\fourteensl
  \fi
  \textfont\ttfam=\fourteentt\def\tt{\fam\ttfam\fourteentt}%
  \ifprod@font
    \scriptfont\ttfam=\tentt
      \scriptscriptfont\ttfam=\seventt
  \else
    \scriptfont\ttfam=\fourteentt
      \scriptscriptfont\ttfam=\fourteentt
  \fi
  \textfont\scfam=\fourteencsc\def\sc{\fam\scfam\fourteencsc}%
  \ifprod@font
    \scriptfont\scfam=\tencsc
      \scriptscriptfont\scfam=\sevencsc
  \else
    \scriptfont\scfam=\fourteencsc
      \scriptscriptfont\scfam=\fourteencsc
  \fi
  \textfont\sffam=\fourteensf\def\sf{\fam\sffam\fourteensf}%
  \ifprod@font
    \scriptfont\sffam=\tensf
      \scriptscriptfont\sffam=\sevensf
  \else
    \scriptfont\sffam=\fourteensf
      \scriptscriptfont\sffam=\fourteensf
  \fi
  \textfont\mibfam=\fourteenmib
    \scriptfont\mibfam=\tenmib
      \scriptscriptfont\mibfam=\sevenmib
  \textfont\sybfam=\fourteensyb
    \scriptfont\sybfam=\tensyb
      \scriptscriptfont\sybfam=\sevensyb
  \ifprod@font
    \textfont\xmfam=\fourteenxm
      \scriptfont\xmfam=\tenxm
        \scriptscriptfont\xmfam=\sevenxm
   \textfont\ymfam=\fourteenym
      \scriptfont\ymfam=\tenym
        \scriptscriptfont\ymfam=\sevenym
  \fi
  \def\oldstyle{\fam\@ne\fourteeni}%
  \def\boldstyle{\fam\mibfam\fourteenmib}%
  \b@ls{17pt}\rm%
}

\def\seventeenpoint{
  \def\rm{\fam0\seventeenrm}%
  \textfont0\seventeenrm  \scriptfont0\twelverm  \scriptscriptfont0\tenrm%
  \textfont1\seventeeni   \scriptfont1\twelvei   \scriptscriptfont1\teni%
  \textfont2\seventeensy  \scriptfont2\twelvesy  \scriptscriptfont2\tensy%
  \textfont\itfam=\seventeenit\def\it{\fam\itfam\seventeenit}%
  \ifprod@font
    \scriptfont\itfam=\twelveit
      \scriptscriptfont\itfam=\tenit
  \else
    \scriptfont\itfam=\seventeenit
      \scriptscriptfont\itfam=\seventeenit
  \fi
  \textfont\bffam=\seventeenbf%
    \scriptfont\bffam=\twelvebf%
      \scriptscriptfont\bffam=\tenbf%
  \def\bf{\fam\bffam\seventeenbf}%
  \textfont\slfam=\seventeensl\def\sl{\fam\slfam\seventeensl}%
  \ifprod@font
    \scriptfont\slfam=\twelvesl
      \scriptscriptfont\slfam=\tensl
  \else
    \scriptfont\slfam=\seventeensl
      \scriptscriptfont\slfam=\seventeensl
  \fi
  \textfont\ttfam=\seventeentt\def\tt{\fam\ttfam\seventeentt}%
  \ifprod@font
    \scriptfont\ttfam=\twelvett
      \scriptscriptfont\ttfam=\tentt
  \else
    \scriptfont\ttfam=\seventeentt
      \scriptscriptfont\ttfam=\seventeentt
  \fi
  \textfont\scfam=\seventeencsc\def\sc{\fam\scfam\seventeencsc}%
  \ifprod@font
    \scriptfont\scfam=\twelvecsc
      \scriptscriptfont\scfam=\tencsc
  \else
    \scriptfont\scfam=\seventeencsc
      \scriptscriptfont\scfam=\seventeencsc
  \fi
  \textfont\sffam=\seventeensf\def\sf{\fam\sffam\seventeensf}%
  \ifprod@font
    \scriptfont\sffam=\twelvesf
      \scriptscriptfont\sffam=\tensf
  \else
    \scriptfont\sffam=\seventeensf
      \scriptscriptfont\sffam=\seventeensf
  \fi
  \textfont\mibfam=\seventeenmib
    \scriptfont\mibfam=\twelvemib
      \scriptscriptfont\mibfam=\tenmib
  \textfont\sybfam=\seventeensyb
    \scriptfont\sybfam=\twelvesyb
      \scriptscriptfont\sybfam=\tensyb
  \ifprod@font
    \textfont\xmfam=\seventeenxm
      \scriptfont\xmfam=\twelvexm
        \scriptscriptfont\xmfam=\tenxm
    \textfont\ymfam=\seventeenym
      \scriptfont\ymfam=\twelveym
        \scriptscriptfont\ymfam=\tenym
  \fi
  \def\oldstyle{\fam\@ne\seventeeni}%
  \def\boldstyle{\fam\mibfam\seventeenmib}%
  \b@ls{20pt}\rm%
}

\lineskip=1pt      \normallineskip=\lineskip
\lineskiplimit=\z@ \normallineskiplimit=\lineskiplimit



\def\la{\mathrel{\mathchoice {\vcenter{\offinterlineskip\halign{\hfil
$\displaystyle##$\hfil\cr<\cr\sim\cr}}}
{\vcenter{\offinterlineskip\halign{\hfil$\textstyle##$\hfil\cr
<\cr\sim\cr}}}
{\vcenter{\offinterlineskip\halign{\hfil$\scriptstyle##$\hfil\cr
<\cr\sim\cr}}}
{\vcenter{\offinterlineskip\halign{\hfil$\scriptscriptstyle##$\hfil\cr
<\cr\sim\cr}}}}}

\def\ga{\mathrel{\mathchoice {\vcenter{\offinterlineskip\halign{\hfil
$\displaystyle##$\hfil\cr>\cr\sim\cr}}}
{\vcenter{\offinterlineskip\halign{\hfil$\textstyle##$\hfil\cr
>\cr\sim\cr}}}
{\vcenter{\offinterlineskip\halign{\hfil$\scriptstyle##$\hfil\cr
>\cr\sim\cr}}}
{\vcenter{\offinterlineskip\halign{\hfil$\scriptscriptstyle##$\hfil\cr
>\cr\sim\cr}}}}}

\def\getsto{\mathrel{\mathchoice {\vcenter{\offinterlineskip
\halign{\hfil
$\displaystyle##$\hfil\cr\gets\cr\to\cr}}}
{\vcenter{\offinterlineskip\halign{\hfil$\textstyle##$\hfil\cr\gets
\cr\to\cr}}}
{\vcenter{\offinterlineskip\halign{\hfil$\scriptstyle##$\hfil\cr\gets
\cr\to\cr}}}
{\vcenter{\offinterlineskip\halign{\hfil$\scriptscriptstyle##$\hfil\cr
\gets\cr\to\cr}}}}}

\def\lid{\mathrel{\mathchoice {\vcenter{\offinterlineskip\halign{\hfil
$\displaystyle##$\hfil\cr<\cr\noalign{\vskip1.2pt}=\cr}}}
{\vcenter{\offinterlineskip\halign{\hfil$\textstyle##$\hfil\cr<\cr
\noalign{\vskip1.2pt}=\cr}}}
{\vcenter{\offinterlineskip\halign{\hfil$\scriptstyle##$\hfil\cr<\cr
\noalign{\vskip1pt}=\cr}}}
{\vcenter{\offinterlineskip\halign{\hfil$\scriptscriptstyle##$\hfil\cr
<\cr
\noalign{\vskip0.9pt}=\cr}}}}}

\def\gid{\mathrel{\mathchoice {\vcenter{\offinterlineskip\halign{\hfil
$\displaystyle##$\hfil\cr>\cr\noalign{\vskip1.2pt}=\cr}}}
{\vcenter{\offinterlineskip\halign{\hfil$\textstyle##$\hfil\cr>\cr
\noalign{\vskip1.2pt}=\cr}}}
{\vcenter{\offinterlineskip\halign{\hfil$\scriptstyle##$\hfil\cr>\cr
\noalign{\vskip1pt}=\cr}}}
{\vcenter{\offinterlineskip\halign{\hfil$\scriptscriptstyle##$\hfil\cr
>\cr
\noalign{\vskip0.9pt}=\cr}}}}}

\def\grole{\mathrel{\mathchoice {\vcenter{\offinterlineskip\halign{\hfil
$\displaystyle##$\hfil\cr>\cr\noalign{\vskip-1.5pt}<\cr}}}
{\vcenter{\offinterlineskip\halign{\hfil$\textstyle##$\hfil\cr
>\cr\noalign{\vskip-1.5pt}<\cr}}}
{\vcenter{\offinterlineskip\halign{\hfil$\scriptstyle##$\hfil\cr
>\cr\noalign{\vskip-1pt}<\cr}}}
{\vcenter{\offinterlineskip\halign{\hfil$\scriptscriptstyle##$\hfil\cr
>\cr\noalign{\vskip-0.5pt}<\cr}}}}}

\def\leogr{\mathrel{\mathchoice {\vcenter{\offinterlineskip\halign{\hfil
$\displaystyle##$\hfil\cr<\cr\noalign{\vskip-1.5pt}>\cr}}}
{\vcenter{\offinterlineskip\halign{\hfil$\textstyle##$\hfil\cr
<\cr\noalign{\vskip-1.5pt}>\cr}}}
{\vcenter{\offinterlineskip\halign{\hfil$\scriptstyle##$\hfil\cr
<\cr\noalign{\vskip-1pt}>\cr}}}
{\vcenter{\offinterlineskip\halign{\hfil$\scriptscriptstyle##$\hfil\cr
<\cr\noalign{\vskip-0.5pt}>\cr}}}}}

\def\loa{\mathrel{\mathchoice {\vcenter{\offinterlineskip\halign{\hfil
$\displaystyle##$\hfil\cr<\cr\approx\cr}}}
{\vcenter{\offinterlineskip\halign{\hfil$\textstyle##$\hfil\cr
<\cr\approx\cr}}}
{\vcenter{\offinterlineskip\halign{\hfil$\scriptstyle##$\hfil\cr
<\cr\approx\cr}}}
{\vcenter{\offinterlineskip\halign{\hfil$\scriptscriptstyle##$\hfil\cr
<\cr\approx\cr}}}}}

\def\goa{\mathrel{\mathchoice {\vcenter{\offinterlineskip\halign{\hfil
$\displaystyle##$\hfil\cr>\cr\approx\cr}}}
{\vcenter{\offinterlineskip\halign{\hfil$\textstyle##$\hfil\cr
>\cr\approx\cr}}}
{\vcenter{\offinterlineskip\halign{\hfil$\scriptstyle##$\hfil\cr
>\cr\approx\cr}}}
{\vcenter{\offinterlineskip\halign{\hfil$\scriptscriptstyle##$\hfil\cr
>\cr\approx\cr}}}}}

\def\sun{\hbox{$\odot$}}

\def\diameter{{\ifmmode\mathchoice
{\ooalign{\hfil\hbox{$\displaystyle/$}\hfil\crcr
{\hbox{$\displaystyle\mathchar"20D$}}}}
{\ooalign{\hfil\hbox{$\textstyle/$}\hfil\crcr
{\hbox{$\textstyle\mathchar"20D$}}}}
{\ooalign{\hfil\hbox{$\scriptstyle/$}\hfil\crcr
{\hbox{$\scriptstyle\mathchar"20D$}}}}
{\ooalign{\hfil\hbox{$\scriptscriptstyle/$}\hfil\crcr
{\hbox{$\scriptscriptstyle\mathchar"20D$}}}}
\else{\ooalign{\hfil/\hfil\crcr\mathhexbox20D}}%
\fi}}

\def\sq{\ifmmode\squareforqed\else{\unskip\nobreak\hfil
\penalty50\hskip1em\null\nobreak\hfil\squareforqed
\parfillskip=0pt\finalhyphendemerits=0\endgraf}\fi}
\def\squareforqed{\hbox{\rlap{$\sqcap$}$\sqcup$}}


\def\bbbc{{\mathchoice {\setbox0=\hbox{$\displaystyle\rm C$}\hbox{\hbox
to0pt{\kern0.4\wd0\vrule height0.9\ht0\hss}\box0}}
{\setbox0=\hbox{$\textstyle\rm C$}\hbox{\hbox
to0pt{\kern0.4\wd0\vrule height0.9\ht0\hss}\box0}}
{\setbox0=\hbox{$\scriptstyle\rm C$}\hbox{\hbox
to0pt{\kern0.4\wd0\vrule height0.9\ht0\hss}\box0}}
{\setbox0=\hbox{$\scriptscriptstyle\rm C$}\hbox{\hbox
to0pt{\kern0.4\wd0\vrule height0.9\ht0\hss}\box0}}}}
\def\bbbq{{\mathchoice {\setbox0=\hbox{$\displaystyle\rm
Q$}\hbox{\raise
0.15\ht0\hbox to0pt{\kern0.4\wd0\vrule height0.8\ht0\hss}\box0}}
{\setbox0=\hbox{$\textstyle\rm Q$}\hbox{\raise
0.15\ht0\hbox to0pt{\kern0.4\wd0\vrule height0.8\ht0\hss}\box0}}
{\setbox0=\hbox{$\scriptstyle\rm Q$}\hbox{\raise
0.15\ht0\hbox to0pt{\kern0.4\wd0\vrule height0.7\ht0\hss}\box0}}
{\setbox0=\hbox{$\scriptscriptstyle\rm Q$}\hbox{\raise
0.15\ht0\hbox to0pt{\kern0.4\wd0\vrule height0.7\ht0\hss}\box0}}}}
\def\bbbt{{\mathchoice {\setbox0=\hbox{$\displaystyle\rm
T$}\hbox{\hbox to0pt{\kern0.3\wd0\vrule height0.9\ht0\hss}\box0}}
{\setbox0=\hbox{$\textstyle\rm T$}\hbox{\hbox
to0pt{\kern0.3\wd0\vrule height0.9\ht0\hss}\box0}}
{\setbox0=\hbox{$\scriptstyle\rm T$}\hbox{\hbox
to0pt{\kern0.3\wd0\vrule height0.9\ht0\hss}\box0}}
{\setbox0=\hbox{$\scriptscriptstyle\rm T$}\hbox{\hbox
to0pt{\kern0.3\wd0\vrule height0.9\ht0\hss}\box0}}}}
\def\bbbs{{\mathchoice
{\setbox0=\hbox{$\displaystyle     \rm S$}\hbox{\raise0.5\ht0\hbox
to0pt{\kern0.35\wd0\vrule height0.45\ht0\hss}\hbox
to0pt{\kern0.55\wd0\vrule height0.5\ht0\hss}\box0}}
{\setbox0=\hbox{$\textstyle        \rm S$}\hbox{\raise0.5\ht0\hbox
to0pt{\kern0.35\wd0\vrule height0.45\ht0\hss}\hbox
to0pt{\kern0.55\wd0\vrule height0.5\ht0\hss}\box0}}
{\setbox0=\hbox{$\scriptstyle      \rm S$}\hbox{\raise0.5\ht0\hbox
to0pt{\kern0.35\wd0\vrule height0.45\ht0\hss}\raise0.05\ht0\hbox
to0pt{\kern0.5\wd0\vrule height0.45\ht0\hss}\box0}}
{\setbox0=\hbox{$\scriptscriptstyle\rm S$}\hbox{\raise0.5\ht0\hbox
to0pt{\kern0.4\wd0\vrule height0.45\ht0\hss}\raise0.05\ht0\hbox
to0pt{\kern0.55\wd0\vrule height0.45\ht0\hss}\box0}}}}
\def\bbbz{{\mathchoice {\hbox{$\sf\textstyle Z\kern-0.4em Z$}}
{\hbox{$\sf\textstyle Z\kern-0.4em Z$}}
{\hbox{$\sf\scriptstyle Z\kern-0.3em Z$}}
{\hbox{$\sf\scriptscriptstyle Z\kern-0.2em Z$}}}}


\ifprod@font
  \mathchardef\la="3\@xm2E
  \mathchardef\getsto="3\@xm1C
  \mathchardef\lid="3\@xm35
  \mathchardef\grole="3\@xm3F
  \mathchardef\loa="3\@xm2F
  \mathchardef\ga="3\@xm26
  \mathchardef\gid="3\@xm3D
  \mathchardef\leogr="3\@xm37
  \mathchardef\goa="3\@xm27
  \mathchardef\sq="0\@xm03
%
%
\def\diameter{{%
  \ifmmode
    \mathchoice
    {\ooalign{\hfil\hbox{$\displaystyle/$}\hfil\crcr
    {\lower.2ex\hbox{$\displaystyle\mathchar"20D$}}}}%
    {\ooalign{\hfil\hbox{$\textstyle/$}\hfil\crcr
    {\lower.2ex\hbox{$\textstyle\mathchar"20D$}}}}%
    {\ooalign{\hfil\hbox{$\scriptstyle/$}\hfil\crcr
    {\lower.1ex\hbox{$\scriptstyle\mathchar"20D$}}}}%
    {\ooalign{\hfil\hbox{$\scriptscriptstyle/$}\hfil\crcr
    {\lower.1ex\hbox{$\scriptscriptstyle\mathchar"20D$}}}}%
  \else
    {\ooalign{\hfil/\hfil\crcr\lower.2ex\hbox{\mathhexbox20D}}}%
  \fi
}}
%
%

\def\bbbc{{\Bbb{C}}}
\def\bbbq{{\Bbb{Q}}}
\def\bbbt{{\Bbb{T}}}
\def\bbbs{{\Bbb{S}}}
\def\bbbz{{\Bbb{Z}}}
\fi


\ifprod@font
\mathchardef\boxdot="2\@xm00
\mathchardef\boxplus="2\@xm01
\mathchardef\boxtimes="2\@xm02
\mathchardef\square="0\@xm03
\mathchardef\blacksquare="0\@xm04
\mathchardef\centerdot="2\@xm05
\mathchardef\lozenge="0\@xm06
\mathchardef\blacklozenge="0\@xm07
\mathchardef\circlearrowright="3\@xm08
\mathchardef\circlearrowleft="3\@xm09
\mathchardef\rightleftharpoons="3\@xm0A
\mathchardef\leftrightharpoons="3\@xm0B
\mathchardef\boxminus="2\@xm0C
\mathchardef\Vdash="3\@xm0D
\mathchardef\Vvdash="3\@xm0E
\mathchardef\vDash="3\@xm0F
\mathchardef\twoheadrightarrow="3\@xm10
\mathchardef\twoheadleftarrow="3\@xm11
\mathchardef\leftleftarrows="3\@xm12
\mathchardef\rightrightarrows="3\@xm13
\mathchardef\upuparrows="3\@xm14
\mathchardef\downdownarrows="3\@xm15
\mathchardef\upharpoonright="3\@xm16

\mathchardef\downharpoonright="3\@xm17
\mathchardef\upharpoonleft="3\@xm18
\mathchardef\downharpoonleft="3\@xm19
\mathchardef\rightarrowtail="3\@xm1A
\mathchardef\leftarrowtail="3\@xm1B
\mathchardef\leftrightarrows="3\@xm1C
\mathchardef\rightleftarrows="3\@xm1D
\mathchardef\Lsh="3\@xm1E
\mathchardef\Rsh="3\@xm1F
\mathchardef\rightsquigarrow="3\@xm20
\mathchardef\leftrightsquigarrow="3\@xm21
\mathchardef\looparrowleft="3\@xm22
\mathchardef\looparrowright="3\@xm23
\mathchardef\circeq="3\@xm24
\mathchardef\succsim="3\@xm25
\mathchardef\gtrsim="3\@xm26
\mathchardef\gtrapprox="3\@xm27
\mathchardef\multimap="3\@xm28
\mathchardef\therefore="3\@xm29
\mathchardef\because="3\@xm2A
\mathchardef\doteqdot="3\@xm2B

\mathchardef\triangleq="3\@xm2C
\mathchardef\precsim="3\@xm2D
\mathchardef\lesssim="3\@xm2E
\mathchardef\lessapprox="3\@xm2F
\mathchardef\eqslantless="3\@xm30
\mathchardef\eqslantgtr="3\@xm31
\mathchardef\curlyeqprec="3\@xm32
\mathchardef\curlyeqsucc="3\@xm33
\mathchardef\preccurlyeq="3\@xm34
\mathchardef\leqq="3\@xm35
\mathchardef\leqslant="3\@xm36
\mathchardef\lessgtr="3\@xm37
\mathchardef\backprime="0\@xm38
\mathchardef\risingdotseq="3\@xm3A
\mathchardef\fallingdotseq="3\@xm3B
\mathchardef\succcurlyeq="3\@xm3C
\mathchardef\geqq="3\@xm3D
\mathchardef\geqslant="3\@xm3E
\mathchardef\gtrless="3\@xm3F
\mathchardef\sqsubset="3\@xm40
\mathchardef\sqsupset="3\@xm41
\mathchardef\vartriangleright="3\@xm42
\mathchardef\vartriangleleft="3\@xm43
\mathchardef\trianglerighteq="3\@xm44
\mathchardef\trianglelefteq="3\@xm45
\mathchardef\bigstar="0\@xm46
\mathchardef\between="3\@xm47
\mathchardef\blacktriangledown="0\@xm48
\mathchardef\blacktriangleright="3\@xm49
\mathchardef\blacktriangleleft="3\@xm4A
\mathchardef\vartriangle="0\@xm4D
\mathchardef\blacktriangle="0\@xm4E
\mathchardef\triangledown="0\@xm4F
\mathchardef\eqcirc="3\@xm50
\mathchardef\lesseqgtr="3\@xm51
\mathchardef\gtreqless="3\@xm52
\mathchardef\lesseqqgtr="3\@xm53
\mathchardef\gtreqqless="3\@xm54
\mathchardef\Rrightarrow="3\@xm56
\mathchardef\Lleftarrow="3\@xm57
\mathchardef\veebar="2\@xm59
\mathchardef\barwedge="2\@xm5A
\mathchardef\doublebarwedge="2\@xm5B
\mathchardef\angle="0\@xm5C
\mathchardef\measuredangle="0\@xm5D
\mathchardef\sphericalangle="0\@xm5E
\mathchardef\varpropto="3\@xm5F
\mathchardef\smallsmile="3\@xm60
\mathchardef\smallfrown="3\@xm61
\mathchardef\Subset="3\@xm62
\mathchardef\Supset="3\@xm63
\mathchardef\Cup="2\@xm64

\mathchardef\Cap="2\@xm65

\mathchardef\curlywedge="2\@xm66
\mathchardef\curlyvee="2\@xm67
\mathchardef\leftthreetimes="2\@xm68
\mathchardef\rightthreetimes="2\@xm69
\mathchardef\subseteqq="3\@xm6A
\mathchardef\supseteqq="3\@xm6B
\mathchardef\bumpeq="3\@xm6C
\mathchardef\Bumpeq="3\@xm6D
\mathchardef\lll="3\@xm6E

\mathchardef\ggg="3\@xm6F

\mathchardef\circledS="0\@xm73
\mathchardef\pitchfork="3\@xm74
\mathchardef\dotplus="2\@xm75
\mathchardef\backsim="3\@xm76
\mathchardef\backsimeq="3\@xm77
\mathchardef\complement="0\@xm7B
\mathchardef\intercal="2\@xm7C
\mathchardef\circledcirc="2\@xm7D
\mathchardef\circledast="2\@xm7E
\mathchardef\circleddash="2\@xm7F
\def\ulcorner{\delimiter"4\@xm70\@xm70 }
\def\urcorner{\delimiter"5\@xm71\@xm71 }
\def\llcorner{\delimiter"4\@xm78\@xm78 }
\def\lrcorner{\delimiter"5\@xm79\@xm79 }
\def\yen{\mathhexbox\@xm55 }
\def\checkmark{\mathhexbox\@xm58 }
\def\circledR{\mathhexbox\@xm72 }
\def\maltese{\mathhexbox\@xm7A }
\mathchardef\lvertneqq="3\@ym00
\mathchardef\gvertneqq="3\@ym01
\mathchardef\nleq="3\@ym02
\mathchardef\ngeq="3\@ym03
\mathchardef\nless="3\@ym04
\mathchardef\ngtr="3\@ym05
\mathchardef\nprec="3\@ym06
\mathchardef\nsucc="3\@ym07
\mathchardef\lneqq="3\@ym08
\mathchardef\gneqq="3\@ym09
\mathchardef\nleqslant="3\@ym0A
\mathchardef\ngeqslant="3\@ym0B
\mathchardef\lneq="3\@ym0C
\mathchardef\gneq="3\@ym0D
\mathchardef\npreceq="3\@ym0E
\mathchardef\nsucceq="3\@ym0F
\mathchardef\precnsim="3\@ym10
\mathchardef\succnsim="3\@ym11
\mathchardef\lnsim="3\@ym12
\mathchardef\gnsim="3\@ym13
\mathchardef\nleqq="3\@ym14
\mathchardef\ngeqq="3\@ym15
\mathchardef\precneqq="3\@ym16
\mathchardef\succneqq="3\@ym17
\mathchardef\precnapprox="3\@ym18
\mathchardef\succnapprox="3\@ym19
\mathchardef\lnapprox="3\@ym1A
\mathchardef\gnapprox="3\@ym1B
\mathchardef\nsim="3\@ym1C
\mathchardef\ncong="3\@ym1D

\mathchardef\varsubsetneq="3\@ym20
\mathchardef\varsupsetneq="3\@ym21
\mathchardef\nsubseteqq="3\@ym22
\mathchardef\nsupseteqq="3\@ym23
\mathchardef\subsetneqq="3\@ym24
\mathchardef\supsetneqq="3\@ym25
\mathchardef\varsubsetneqq="3\@ym26
\mathchardef\varsupsetneqq="3\@ym27
\mathchardef\subsetneq="3\@ym28
\mathchardef\supsetneq="3\@ym29
\mathchardef\nsubseteq="3\@ym2A
\mathchardef\nsupseteq="3\@ym2B
\mathchardef\nparallel="3\@ym2C
\mathchardef\nmid="3\@ym2D
\mathchardef\nshortmid="3\@ym2E
\mathchardef\nshortparallel="3\@ym2F
\mathchardef\nvdash="3\@ym30
\mathchardef\nVdash="3\@ym31
\mathchardef\nvDash="3\@ym32
\mathchardef\nVDash="3\@ym33
\mathchardef\ntrianglerighteq="3\@ym34
\mathchardef\ntrianglelefteq="3\@ym35
\mathchardef\ntriangleleft="3\@ym36
\mathchardef\ntriangleright="3\@ym37
\mathchardef\nleftarrow="3\@ym38
\mathchardef\nrightarrow="3\@ym39
\mathchardef\nLeftarrow="3\@ym3A
\mathchardef\nRightarrow="3\@ym3B
\mathchardef\nLeftrightarrow="3\@ym3C
\mathchardef\nleftrightarrow="3\@ym3D
\mathchardef\divideontimes="2\@ym3E
\mathchardef\varnothing="0\@ym3F
\mathchardef\nexists="0\@ym40
\mathchardef\mho="0\@ym66
\mathchardef\eth="0\@ym67
\mathchardef\eqsim="3\@ym68
\mathchardef\beth="0\@ym69
\mathchardef\gimel="0\@ym6A
\mathchardef\daleth="0\@ym6B
\mathchardef\lessdot="3\@ym6C
\mathchardef\gtrdot="3\@ym6D
\mathchardef\ltimes="2\@ym6E
\mathchardef\rtimes="2\@ym6F
\mathchardef\shortmid="3\@ym70
\mathchardef\shortparallel="3\@ym71
\mathchardef\smallsetminus="2\@ym72
\mathchardef\thicksim="3\@ym73
\mathchardef\thickapprox="3\@ym74
\mathchardef\approxeq="3\@ym75
\mathchardef\succapprox="3\@ym76
\mathchardef\precapprox="3\@ym77
\mathchardef\curvearrowleft="3\@ym78
\mathchardef\curvearrowright="3\@ym79
\mathchardef\digamma="0\@ym7A
\mathchardef\varkappa="0\@ym7B
\mathchardef\hslash="0\@ym7D
\mathchardef\hbar="0\@ym7E
\mathchardef\backepsilon="3\@ym7F


\def\Bbb{\ifmmode\let\next\Bbb@\else
\def\next{\errmessage{Use \string\Bbb\space only in math mode}}\fi\next}
\def\Bbb@#1{{\Bbb@@{#1}}}
\def\Bbb@@#1{\fam\ymfam#1}
\fi


\def\Nulle{0} 
\def\Afe{1}   
\def\Hae{2}   
\def\Hbe{3}   
\def\Hce{4}   
\def\Hde{5}   


\newcount\LastMac       \LastMac=\Nulle

\newskip\half      \half=5.5pt plus 1.5pt minus 2.25pt
\newskip\one       \one=11pt plus 3pt minus 5.5pt
\newskip\onehalf   \onehalf=16.5pt plus 5.5pt minus 8.25pt
\newskip\two       \two=22pt plus 5.5pt minus 11pt

\def\Half{\addvspace{\half}}
\def\One{\addvspace{\one}}
\def\OneHalf{\addvspace{\onehalf}}
\def\Two{\addvspace{\two}}


\def\Raggedright{
  \rightskip=\z@ plus \hsize\relax
}

\def\Fullout{
  \rightskip=\z@\relax
}

\def\Hang#1#2{
  \hangindent=#1%
  \hangafter=#2\relax
}


\newif\ifsp@page
\def\pagestyle#1{\csname ps@#1\endcsname}
\def\thispagestyle#1{\global\sp@pagetrue\gdef\sp@type{#1}}

\def\ps@titlepage{%
  \def\@oddhead{\eightpoint\noindent \the\CatchLine
    \ifprod@font\else\qquad Printed\ \today\fi \hfil}%
  \let\@evenhead=\@oddhead
}

\def\ps@headings{%
  \def\@oddhead{\elevenpoint\it\noindent
    \hfill\the\RightHeader\hskip1.5em\rm\folio}%
  \def\@evenhead{\elevenpoint\noindent
    \folio\hskip1.5em\it\the\LeftHeader\hfill}%
}

\def\ps@plate{%
  \def\@oddhead{\eightpoint\noindent\plt@cap\hfil}%
  \def\@evenhead{\eightpoint\noindent\plt@cap\hfil}%
}



\def\title#1{
  \bgroup
    \vbox to 8pt{\vss}%
    \seventeenpoint
    \Raggedright
    \noindent \strut{\bf #1}\par
  \egroup
}

\def\author#1{
  \bgroup
    \ifnum\LastMac=\Afe \OneHalf\else \vskip 21pt\fi
    \fourteenpoint
    \Raggedright
    \noindent \strut #1\par
    \vskip 3pt%
  \egroup
}

\def\affiliation#1{
  \bgroup
    \vskip -4pt%
    \eightpoint
    \Raggedright
    \noindent \strut {\it #1}\par
  \egroup
  \LastMac=\Afe\relax
}

\def\acceptedline#1{
  \bgroup
    \Two
    \eightpoint
    \Raggedright
    \noindent \strut #1\par
  \egroup
}

\long\def\abstract#1{%
  \bgroup
    \vskip 20pt%
    \everypar{\Hang{11pc}{0}}%
    \noindent{\ninebf ABSTRACT}\par
    \tenpoint
    \Fullout
    \noindent #1\par
  \egroup
}

\long\def\keywords#1{
  \bgroup
    \Half
    \everypar{\Hang{11pc}{0}}%
    \tenpoint
    \Fullout
    \noindent\hbox{\bf Key words:}\ #1\par
  \egroup
}


\def\maketitle{%
  \EndOpening
  \ifsinglecol \else \MakePage\fi
}



\def\Autonumber{
  \global\AutoNumbertrue  
}

\newif\ifAutoNumber \AutoNumberfalse
\newcount\Sec        
\newcount\SecSec
\newcount\SecSecSec

\Sec=\z@

\def\:{\let\@sptoken= } \:  
\def\:{\@xifnch} \expandafter\def\: {\futurelet\@tempc\@ifnch}

\def\@ifnextchar#1#2#3{%
  \let\@tempMACe #1%
  \def\@tempMACa{#2}%
  \def\@tempMACb{#3}%
  \futurelet \@tempMACc\@ifnch%
}

\def\@ifnch{%
\ifx \@tempMACc \@sptoken%
  \let\@tempMACd\@xifnch%
\else%
  \ifx \@tempMACc \@tempMACe%
    \let\@tempMACd\@tempMACa%
  \else%
    \let\@tempMACd\@tempMACb%
  \fi%
\fi%
\@tempMACd%
}

\def\@ifstar#1#2{\@ifnextchar *{\def\@tempMACa*{#1}\@tempMACa}{#2}}

\newskip\@tempskipb

\def\addvspace#1{%
  \ifvmode\else \endgraf\fi%
  \ifdim\lastskip=\z@%
    \vskip #1\relax%
  \else%
    \@tempskipb#1\relax\@xaddvskip%
  \fi%
}

\def\@xaddvskip{%
  \ifdim\lastskip<\@tempskipb%
    \vskip-\lastskip%
    \vskip\@tempskipb\relax%
  \else%
    \ifdim\@tempskipb<\z@%
      \ifdim\lastskip<\z@ \else%
        \advance\@tempskipb\lastskip%
        \vskip-\lastskip\vskip\@tempskipb%
      \fi%
    \fi%
  \fi%
}

\newskip\@tmpSKIP

\def\addpen#1{%
  \ifvmode
    \if@nobreak
    \else
      \ifdim\lastskip=\z@
        \penalty#1\relax
      \else
        \@tmpSKIP=\lastskip
        \vskip -\lastskip
        \penalty#1\vskip\@tmpSKIP
      \fi
    \fi
  \fi
}

\newcount\@clubpen   \@clubpen=\clubpenalty
\newif\if@nobreak    \@nobreakfalse

\def\@noafterindent{%
  \global\@nobreaktrue
  \everypar{\if@nobreak
              \global\@nobreakfalse
              \clubpenalty \@M
              {\setbox\z@\lastbox}%
              \LastMac=\Nulle\relax%
            \else
              \clubpenalty \@clubpen
              \everypar{}%
            \fi}
}

\newcount\gds@cbrk   \gds@cbrk=-300

\def\@nohdbrk{\interlinepenalty \@M\relax}

\let\@par=\par
\def\@restorepar{\def\par{\@par}}

\newif\if@endpe   \@endpefalse
 
\def\@doendpe{\@endpetrue \@nobreakfalse \LastMac=\Nulle\relax%
     \def\par{\@restorepar\everypar{}\par\@endpefalse}%
              \everypar{\setbox\z@\lastbox\everypar{}\@endpefalse}%
}

\def\section{\@ifstar{\@ssection}{\@section}}

\def\@section#1{
  \if@nobreak
    \everypar{}%
    \ifnum\LastMac=\Hae \addvspace{\half}\fi
  \else
    \addpen{\gds@cbrk}%
    \addvspace{\two}%
  \fi
  \bgroup
    \ninepoint\bf
    \Raggedright
    \ifAutoNumber
      \global\advance\Sec \@ne
      \noindent\@nohdbrk\number\Sec\hskip 1pc \uppercase{#1}\par
      \global\SecSec=\z@
    \else
      \noindent\@nohdbrk\uppercase{#1}\par
    \fi
  \egroup
  \nobreak
  \vskip\half
  \nobreak
  \@noafterindent
  \LastMac=\Hae\relax
}

\def\@ssection#1{
  \if@nobreak
    \everypar{}%
    \ifnum\LastMac=\Hae \addvspace{\half}\fi
  \else
    \addpen{\gds@cbrk}%
    \addvspace{\two}%
  \fi
  \bgroup
    \ninepoint\bf
    \Raggedright
    \noindent\@nohdbrk\uppercase{#1}\par
  \egroup
  \nobreak
  \vskip\half
  \nobreak
  \@noafterindent
  \LastMac=\Hae\relax
}

\def\subsection#1{
  \if@nobreak
    \everypar{}%
    \ifnum\LastMac=\Hae \addvspace{1pt plus 1pt minus .5pt}\fi
  \else
    \addpen{\gds@cbrk}%
    \addvspace{\onehalf}%
  \fi
  \bgroup
    \ninepoint\bf
    \Raggedright
    \ifAutoNumber
      \global\advance\SecSec \@ne
      \noindent\@nohdbrk\number\Sec.\number\SecSec \hskip 1pc\relax #1\par
      \global\SecSecSec=\z@
    \else
      \noindent\@nohdbrk #1\par
    \fi
  \egroup
  \nobreak
  \vskip\half
  \nobreak
  \@noafterindent
  \LastMac=\Hbe\relax
}

\def\subsubsection#1{
  \if@nobreak
    \everypar{}%
    \ifnum\LastMac=\Hbe \addvspace{1pt plus 1pt minus .5pt}\fi
  \else
    \addpen{\gds@cbrk}%
    \addvspace{\onehalf}%
  \fi
  \bgroup
    \ninepoint\it
    \Raggedright
    \ifAutoNumber
      \global\advance\SecSecSec \@ne
      \noindent\@nohdbrk\number\Sec.\number\SecSec.\number\SecSecSec
        \hskip 1pc\relax #1\par
    \else
      \noindent\@nohdbrk #1\par
    \fi
  \egroup
  \nobreak
  \vskip\half
  \nobreak
  \@noafterindent
  \LastMac=\Hce\relax
}

\def\paragraph#1{
  \if@nobreak
    \everypar{}%
  \else
    \addpen{\gds@cbrk}%
    \addvspace{\one}%
  \fi%
  \bgroup%
    \ninepoint\it
    \noindent #1\ \nobreak%
  \egroup
  \LastMac=\Hde\relax
  \ignorespaces
}




\def\beginlist{%
  \par\if@nobreak \else\addvspace{\half}\fi%
  \bgroup%
    \ninepoint
    \let\item=\list@item%
}

\def\list@item{%
  \par\noindent\hskip 1em\relax%
  \ignorespaces%
}

\def\endlist{\par\egroup\addvspace{\half}\@doendpe}


\def\beginrefs{%
  \par
  \bgroup
    \eightpoint
    \Raggedright
    \let\bibitem=\bib@item
}

\def\bib@item{%
  \par\parindent=1.5em\Hang{1.5em}{1}%
  \everypar={\Hang{1.5em}{1}\ignorespaces}%
  \noindent\ignorespaces
}

\def\endrefs{\par\egroup\@doendpe}


\newtoks\CatchLine

\def\@journal{Mon.\ Not.\ R.\ Astron.\ Soc.\ }  
\def\@pubyear{1996}        
\def\@pagerange{000--000}  
\def\@volume{000}          
\def\@microfiche{}         %

\def\pubyear#1{\gdef\@pubyear{#1}\@makecatchline}
\def\pagerange#1{\gdef\@pagerange{#1}\@makecatchline}
\def\volume#1{\gdef\@volume{#1}\@makecatchline}
\def\microfiche#1{\gdef\@microfiche{and Microfiche\ #1}\@makecatchline}

\def\@makecatchline{%
  \global\CatchLine{%
    {\rm \@journal {\bf \@volume},\ \@pagerange\ (\@pubyear)\ \@microfiche}}%
}

\@makecatchline 

\newtoks\LeftHeader
\def\shortauthor#1{
  \global\LeftHeader{#1}%
}

\newtoks\RightHeader
\def\shorttitle#1{
  \global\RightHeader{#1}%
}

\def\PageHead{
  \begingroup
    \ifsp@page
      \csname ps@\sp@type\endcsname
      \global\sp@pagefalse
    \fi
    \ifodd\pageno
      \let\the@head=\@oddhead
    \else
      \let\the@head=\@evenhead
    \fi
    \vbox to \z@{\vskip-22.5\p@%
      \hbox to \PageWidth{\vbox to8.5\p@{}%
        \the@head
      }%
    \vss}%
  \endgroup
  \nointerlineskip
}

\def\today{%
  \number\day\space
  \ifcase\month\or January\or February\or March\or April\or May\or June\or
    July\or August\or September\or October\or November\or December\fi
  \space\number\year%
}

\def\PageFoot{} 

\def\authorcomment#1{%
  \gdef\PageFoot{%
    \nointerlineskip%
    \vbox to 22pt{\vfil%
      \hbox to \PageWidth{\elevenpoint\noindent \hfil #1 \hfil}}%
  }%
}


\newif\ifplate@page
\newbox\plt@box

\def\beginplatepage{%
  \let\plate=\plate@head
  \let\caption=\fig@caption
  \global\setbox\plt@box=\vbox\bgroup
  \TEMPDIMEN=\PageWidth 
  \hsize=\PageWidth\relax
}

\def\endplatepage{\par\egroup\global\plate@pagetrue}
\def\plate@head#1{\gdef\plt@cap{#1}}


\def\letters{%
  \gdef\folio{\ifnum\pageno<\z@ L\romannumeral-\pageno
    \else L\number\pageno \fi}%
}


\everydisplay{\displaysetup}

\newif\ifeqno
\newif\ifleqno

\def\displaysetup#1$${%
 \displaytest#1\eqno\eqno\displaytest
}

\def\displaytest#1\eqno#2\eqno#3\displaytest{%
 \if!#3!\ldisplaytest#1\leqno\leqno\ldisplaytest
 \else\eqnotrue\leqnofalse\def\eqn{#2}\def\eq{#1}\fi
 \generaldisplay$$}

\def\ldisplaytest#1\leqno#2\leqno#3\ldisplaytest{%
 \def\eq{#1}%
 \if!#3!\eqnofalse\else\eqnotrue\leqnotrue
  \def\eqn{#2}\fi}

\def\generaldisplay{%
\ifeqno \ifleqno 
   \hbox to \hsize{\noindent
     $\displaystyle\eq$\hfil$\displaystyle\eqn$}
  \else
    \hbox to \hsize{\noindent
     $\displaystyle\eq$\hfil$\displaystyle\eqn$}
  \fi
 \else
 \hbox to \hsize{\vbox{\noindent
  $\displaystyle\eq$\hfil}}
 \fi
}


\def\@notice{%
  \par\Two%
  \noindent{\b@ls{11pt}\ninerm This paper has been produced using the
    Blackwell Scientific Publications \TeX\ macros.\par}%
}

\outer\def\bye{\@notice\par\vfill\supereject\end}


\def\start@mess{%
  Monthly notices of the RAS journal style (\@typeface)\space
    v\@version,\space \@verdate.%
}

\everyjob{\Warn{\start@mess}}



\newif\if@debug \@debugfalse  

\def\Print#1{\if@debug\immediate\write16{#1}\else \fi}
\def\Warn#1{\immediate\write16{#1}}
\def\wlog#1{}

\newcount\Iteration 

\def\Single{0} \def\Double{1}                 
\def\Figure{0} \def\Table{1}                  

\def\InStack{0}  
\def\InZoneA{1}
\def\InZoneB{2}
\def\InZoneC{3}

\newcount\TEMPCOUNT 
\newdimen\TEMPDIMEN 
\newbox\TEMPBOX     
\newbox\VOIDBOX     

\newcount\LengthOfStack 
\newcount\MaxItems      
\newcount\StackPointer
\newcount\Point         
\newcount\NextFigure    
\newcount\NextTable     
\newcount\NextItem      

\newcount\StatusStack   
\newcount\NumStack      
\newcount\TypeStack     
\newcount\SpanStack     
\newcount\BoxStack      

\newcount\ItemSTATUS    
\newcount\ItemNUMBER    
\newcount\ItemTYPE      
\newcount\ItemSPAN      
\newbox\ItemBOX         
\newdimen\ItemSIZE      

\newdimen\PageHeight    
\newdimen\TextLeading   
\newdimen\Feathering    
\newcount\LinesPerPage  
\newdimen\ColumnWidth   
\newdimen\ColumnGap     
\newdimen\PageWidth     
\newdimen\BodgeHeight   
\newcount\Leading       

\newdimen\ZoneBSize  
\newdimen\TextSize   
\newbox\ZoneABOX     
\newbox\ZoneBBOX     
\newbox\ZoneCBOX     

\newif\ifFirstSingleItem
\newif\ifFirstZoneA
\newif\ifMakePageInComplete
\newif\ifMoreFigures \MoreFiguresfalse 
\newif\ifMoreTables  \MoreTablesfalse  

\newif\ifFigInZoneB 
\newif\ifFigInZoneC 
\newif\ifTabInZoneB 
\newif\ifTabInZoneC

\newif\ifZoneAFullPage

\newbox\MidBOX    
\newbox\LeftBOX
\newbox\RightBOX
\newbox\PageBOX   

\newif\ifLeftCOL  
\LeftCOLtrue

\newdimen\ZoneBAdjust

\newcount\ItemFits
\def\Yes{1}
\def\No{2}


\MaxItems=15
\NextFigure=\z@        
\NextTable=\@ne

\BodgeHeight=6pt
\TextLeading=11pt    
\Leading=11
\Feathering=\z@      
\LinesPerPage=61     
\topskip=\TextLeading
\ColumnWidth=20pc    
\ColumnGap=2pc       

\newskip\ItemSepamount  
\ItemSepamount=\TextLeading plus \TextLeading minus 4pt

\parskip=\z@ plus .1pt
\parindent=18pt
\widowpenalty=\z@
\clubpenalty=10000
\tolerance=1500
\hbadness=1500
\abovedisplayskip=6pt plus 2pt minus 2pt
\belowdisplayskip=6pt plus 2pt minus 2pt
\abovedisplayshortskip=6pt plus 2pt minus 2pt
\belowdisplayshortskip=6pt plus 2pt minus 2pt

\ninepoint 


\PageHeight=682pt

\PageWidth=2\ColumnWidth
\advance\PageWidth by \ColumnGap

\pagestyle{headings}




\newcount\DUMMY \StatusStack=\allocationnumber
\newcount\DUMMY \newcount\DUMMY \newcount\DUMMY 
\newcount\DUMMY \newcount\DUMMY \newcount\DUMMY 
\newcount\DUMMY \newcount\DUMMY \newcount\DUMMY
\newcount\DUMMY \newcount\DUMMY \newcount\DUMMY 
\newcount\DUMMY \newcount\DUMMY \newcount\DUMMY

\newcount\DUMMY \NumStack=\allocationnumber
\newcount\DUMMY \newcount\DUMMY \newcount\DUMMY 
\newcount\DUMMY \newcount\DUMMY \newcount\DUMMY 
\newcount\DUMMY \newcount\DUMMY \newcount\DUMMY 
\newcount\DUMMY \newcount\DUMMY \newcount\DUMMY 
\newcount\DUMMY \newcount\DUMMY \newcount\DUMMY

\newcount\DUMMY \TypeStack=\allocationnumber
\newcount\DUMMY \newcount\DUMMY \newcount\DUMMY 
\newcount\DUMMY \newcount\DUMMY \newcount\DUMMY 
\newcount\DUMMY \newcount\DUMMY \newcount\DUMMY 
\newcount\DUMMY \newcount\DUMMY \newcount\DUMMY 
\newcount\DUMMY \newcount\DUMMY \newcount\DUMMY

\newcount\DUMMY \SpanStack=\allocationnumber
\newcount\DUMMY \newcount\DUMMY \newcount\DUMMY 
\newcount\DUMMY \newcount\DUMMY \newcount\DUMMY 
\newcount\DUMMY \newcount\DUMMY \newcount\DUMMY 
\newcount\DUMMY \newcount\DUMMY \newcount\DUMMY 
\newcount\DUMMY \newcount\DUMMY \newcount\DUMMY

\newbox\DUMMY   \BoxStack=\allocationnumber
\newbox\DUMMY   \newbox\DUMMY \newbox\DUMMY 
\newbox\DUMMY   \newbox\DUMMY \newbox\DUMMY 
\newbox\DUMMY   \newbox\DUMMY \newbox\DUMMY 
\newbox\DUMMY   \newbox\DUMMY \newbox\DUMMY 
\newbox\DUMMY   \newbox\DUMMY \newbox\DUMMY

\def\wlog{\immediate\write\m@ne}


\def\GetItemAll#1{%
 \GetItemSTATUS{#1}
 \GetItemNUMBER{#1}
 \GetItemTYPE{#1}
 \GetItemSPAN{#1}
 \GetItemBOX{#1}
}

\def\GetItemSTATUS#1{%
 \Point=\StatusStack
 \advance\Point by #1
 \global\ItemSTATUS=\count\Point
}

\def\GetItemNUMBER#1{%
 \Point=\NumStack
 \advance\Point by #1
 \global\ItemNUMBER=\count\Point
}

\def\GetItemTYPE#1{%
 \Point=\TypeStack
 \advance\Point by #1
 \global\ItemTYPE=\count\Point
}

\def\GetItemSPAN#1{%
 \Point\SpanStack
 \advance\Point by #1
 \global\ItemSPAN=\count\Point
}

\def\GetItemBOX#1{%
 \Point=\BoxStack
 \advance\Point by #1
 \global\setbox\ItemBOX=\vbox{\copy\Point}
 \global\ItemSIZE=\ht\ItemBOX
 \global\advance\ItemSIZE by \dp\ItemBOX
 \TEMPCOUNT=\ItemSIZE
 \divide\TEMPCOUNT by \Leading
 \divide\TEMPCOUNT by 65536
 \advance\TEMPCOUNT \@ne
 \ItemSIZE=\TEMPCOUNT pt
 \global\multiply\ItemSIZE by \Leading
}


\def\JoinStack{%
 \ifnum\LengthOfStack=\MaxItems 
  \Warn{WARNING: Stack is full...some items will be lost!}
 \else
  \Point=\StatusStack
  \advance\Point by \LengthOfStack
  \global\count\Point=\ItemSTATUS
  \Point=\NumStack
  \advance\Point by \LengthOfStack
  \global\count\Point=\ItemNUMBER
  \Point=\TypeStack
  \advance\Point by \LengthOfStack
  \global\count\Point=\ItemTYPE
  \Point\SpanStack
  \advance\Point by \LengthOfStack
  \global\count\Point=\ItemSPAN
  \Point=\BoxStack
  \advance\Point by \LengthOfStack
  \global\setbox\Point=\vbox{\copy\ItemBOX}
  \global\advance\LengthOfStack \@ne
  \ifnum\ItemTYPE=\Figure 
   \global\MoreFigurestrue
  \else
   \global\MoreTablestrue
  \fi
 \fi
}


\def\LeaveStack#1{%
 {\Iteration=#1
 \loop
 \ifnum\Iteration<\LengthOfStack
  \advance\Iteration \@ne
  \GetItemSTATUS{\Iteration}
   \advance\Point by \m@ne
   \global\count\Point=\ItemSTATUS
  \GetItemNUMBER{\Iteration}
   \advance\Point by \m@ne
   \global\count\Point=\ItemNUMBER
  \GetItemTYPE{\Iteration}
   \advance\Point by \m@ne
   \global\count\Point=\ItemTYPE
  \GetItemSPAN{\Iteration}
   \advance\Point by \m@ne
   \global\count\Point=\ItemSPAN
  \GetItemBOX{\Iteration}
   \advance\Point by \m@ne
   \global\setbox\Point=\vbox{\copy\ItemBOX}
 \repeat}
 \global\advance\LengthOfStack by \m@ne
}


\newif\ifStackNotClean

\def\CleanStack{%
 \StackNotCleantrue
 {\Iteration=\z@
  \loop
   \ifStackNotClean
    \GetItemSTATUS{\Iteration}
    \ifnum\ItemSTATUS=\InStack
     \advance\Iteration \@ne
     \else
      \LeaveStack{\Iteration}
    \fi
   \ifnum\LengthOfStack<\Iteration
    \StackNotCleanfalse
   \fi
 \repeat}
}


\def\FindItem#1#2{%
 \global\StackPointer=\m@ne 
 {\Iteration=\z@
  \loop
  \ifnum\Iteration<\LengthOfStack
   \GetItemSTATUS{\Iteration}
   \ifnum\ItemSTATUS=\InStack
    \GetItemTYPE{\Iteration}
    \ifnum\ItemTYPE=#1
     \GetItemNUMBER{\Iteration}
     \ifnum\ItemNUMBER=#2
      \global\StackPointer=\Iteration
      \Iteration=\LengthOfStack 
     \fi
    \fi
   \fi
  \advance\Iteration \@ne
 \repeat}
}


\def\FindNext{%
 \global\StackPointer=\m@ne 
 {\Iteration=\z@
  \loop
  \ifnum\Iteration<\LengthOfStack
   \GetItemSTATUS{\Iteration}
   \ifnum\ItemSTATUS=\InStack
    \GetItemTYPE{\Iteration}
   \ifnum\ItemTYPE=\Figure
    \ifMoreFigures
      \global\NextItem=\Figure
      \global\StackPointer=\Iteration
      \Iteration=\LengthOfStack 
    \fi
   \fi
   \ifnum\ItemTYPE=\Table
    \ifMoreTables
      \global\NextItem=\Table
      \global\StackPointer=\Iteration
      \Iteration=\LengthOfStack 
    \fi
   \fi
  \fi
  \advance\Iteration \@ne
 \repeat}
}


\def\ChangeStatus#1#2{%
 \Point=\StatusStack
 \advance\Point by #1
 \global\count\Point=#2
}



\def\Zone{\InZoneA}

\ZoneBAdjust=\z@

\def\MakePage{
 \global\ZoneBSize=\PageHeight
 \global\TextSize=\ZoneBSize
 \global\ZoneAFullPagefalse
 \global\topskip=\TextLeading
 \MakePageInCompletetrue
 \MoreFigurestrue
 \MoreTablestrue
 \FigInZoneBfalse
 \FigInZoneCfalse
 \TabInZoneBfalse
 \TabInZoneCfalse
 \global\FirstSingleItemtrue
 \global\FirstZoneAtrue
 \global\setbox\ZoneABOX=\box\VOIDBOX
 \global\setbox\ZoneBBOX=\box\VOIDBOX
 \global\setbox\ZoneCBOX=\box\VOIDBOX
 \loop
  \ifMakePageInComplete
 \FindNext
 \ifnum\StackPointer=\m@ne
  \NextItem=\m@ne
  \MoreFiguresfalse
  \MoreTablesfalse
 \fi
 \ifnum\NextItem=\Figure
   \FindItem{\Figure}{\NextFigure}
   \ifnum\StackPointer=\m@ne \global\MoreFiguresfalse
   \else
    \GetItemSPAN{\StackPointer}
    \ifnum\ItemSPAN=\Single \def\Zone{\InZoneB}\relax
     \ifFigInZoneC \global\MoreFiguresfalse\fi
    \else
     \def\Zone{\InZoneA}
     \ifFigInZoneB \def\Zone{\InZoneC}\fi
    \fi
   \fi
   \ifMoreFigures\Print{}\FigureItems\fi
 \fi
\ifnum\NextItem=\Table
   \FindItem{\Table}{\NextTable}
   \ifnum\StackPointer=\m@ne \global\MoreTablesfalse
   \else
    \GetItemSPAN{\StackPointer}
    \ifnum\ItemSPAN=\Single\relax
     \ifTabInZoneC \global\MoreTablesfalse\fi
    \else
     \def\Zone{\InZoneA}
     \ifTabInZoneB \def\Zone{\InZoneC}\fi
    \fi
   \fi
   \ifMoreTables\Print{}\TableItems\fi
 \fi
   \MakePageInCompletefalse 
   \ifMoreFigures\MakePageInCompletetrue\fi
   \ifMoreTables\MakePageInCompletetrue\fi
 \repeat
 \ifZoneAFullPage
  \global\TextSize=\z@
  \global\ZoneBSize=\z@
  \global\vsize=\z@\relax
  \global\topskip=\z@\relax
  \vbox to \z@{\vss}
  \eject
 \else
 \global\advance\ZoneBSize by -\ZoneBAdjust
 \global\vsize=\ZoneBSize
 \global\hsize=\ColumnWidth
 \global\ZoneBAdjust=\z@
 \ifdim\TextSize<23pt
 \Warn{}
 \Warn{* Making column fall short: TextSize=\the\TextSize *}
 \vskip-\lastskip\eject\fi
 \fi
}

\def\MakeRightCol{
 \global\TextSize=\ZoneBSize
 \MakePageInCompletetrue
 \MoreFigurestrue
 \MoreTablestrue
 \global\FirstSingleItemtrue
 \global\setbox\ZoneBBOX=\box\VOIDBOX
 \def\Zone{\InZoneB}
 \loop
  \ifMakePageInComplete
 \FindNext
 \ifnum\StackPointer=\m@ne
  \NextItem=\m@ne
  \MoreFiguresfalse
  \MoreTablesfalse
 \fi
 \ifnum\NextItem=\Figure
   \FindItem{\Figure}{\NextFigure}
   \ifnum\StackPointer=\m@ne \MoreFiguresfalse
   \else
    \GetItemSPAN{\StackPointer}
    \ifnum\ItemSPAN=\Double\relax
     \MoreFiguresfalse\fi
   \fi
   \ifMoreFigures\Print{}\FigureItems\fi
 \fi
 \ifnum\NextItem=\Table
   \FindItem{\Table}{\NextTable}
   \ifnum\StackPointer=\m@ne \MoreTablesfalse
   \else
    \GetItemSPAN{\StackPointer}
    \ifnum\ItemSPAN=\Double\relax
     \MoreTablesfalse\fi
   \fi
   \ifMoreTables\Print{}\TableItems\fi
 \fi
   \MakePageInCompletefalse 
   \ifMoreFigures\MakePageInCompletetrue\fi
   \ifMoreTables\MakePageInCompletetrue\fi
 \repeat
 \ifZoneAFullPage
  \global\TextSize=\z@
  \global\ZoneBSize=\z@
  \global\vsize=\z@\relax
  \global\topskip=\z@\relax
  \vbox to \z@{\vss}
  \eject
 \else
 \global\vsize=\ZoneBSize
 \global\hsize=\ColumnWidth
 \ifdim\TextSize<23pt
 \Warn{}
 \Warn{* Making column fall short: TextSize=\the\TextSize *}
 \vskip-\lastskip\eject\fi
\fi
}

\def\FigureItems{
 \Print{Considering...}
 \ShowItem{\StackPointer}
 \GetItemBOX{\StackPointer} 
 \GetItemSPAN{\StackPointer}
  \CheckFitInZone 
  \ifnum\ItemFits=\Yes
   \ifnum\ItemSPAN=\Single
     \ChangeStatus{\StackPointer}{\InZoneB} 
     \global\FigInZoneBtrue
     \ifFirstSingleItem
      \hbox{}\vskip-\BodgeHeight
     \global\advance\ItemSIZE by \TextLeading
     \fi
     \unvbox\ItemBOX\ItemSep
     \global\FirstSingleItemfalse
     \global\advance\TextSize by -\ItemSIZE
     \global\advance\TextSize by -\TextLeading
   \else
    \ifFirstZoneA
     \global\advance\ItemSIZE by \TextLeading
     \global\FirstZoneAfalse\fi
    \global\advance\TextSize by -\ItemSIZE
    \global\advance\TextSize by -\TextLeading
    \global\advance\ZoneBSize by -\ItemSIZE
    \global\advance\ZoneBSize by -\TextLeading
    \ifFigInZoneB\relax
     \else
     \ifdim\TextSize<3\TextLeading
     \global\ZoneAFullPagetrue
     \fi
    \fi
    \ChangeStatus{\StackPointer}{\Zone}
    \ifnum\Zone=\InZoneC \global\FigInZoneCtrue\fi
  \fi
   \Print{TextSize=\the\TextSize}
   \Print{ZoneBSize=\the\ZoneBSize}
  \global\advance\NextFigure \@ne
   \Print{This figure has been placed.}
  \else
   \Print{No space available for this figure...holding over.}
   \Print{}
   \global\MoreFiguresfalse
  \fi
}

\def\TableItems{
 \Print{Considering...}
 \ShowItem{\StackPointer}
 \GetItemBOX{\StackPointer} 
 \GetItemSPAN{\StackPointer}
  \CheckFitInZone 
  \ifnum\ItemFits=\Yes
   \ifnum\ItemSPAN=\Single
    \ChangeStatus{\StackPointer}{\InZoneB}
     \global\TabInZoneBtrue
     \ifFirstSingleItem
      \hbox{}\vskip-\BodgeHeight
     \global\advance\ItemSIZE by \TextLeading
     \fi
     \unvbox\ItemBOX\ItemSep
     \global\FirstSingleItemfalse
     \global\advance\TextSize by -\ItemSIZE
     \global\advance\TextSize by -\TextLeading
   \else
    \ifFirstZoneA
    \global\advance\ItemSIZE by \TextLeading
    \global\FirstZoneAfalse\fi
    \global\advance\TextSize by -\ItemSIZE
    \global\advance\TextSize by -\TextLeading
    \global\advance\ZoneBSize by -\ItemSIZE
    \global\advance\ZoneBSize by -\TextLeading
    \ifFigInZoneB\relax
     \else
     \ifdim\TextSize<3\TextLeading
     \global\ZoneAFullPagetrue
     \fi
    \fi
    \ChangeStatus{\StackPointer}{\Zone}
    \ifnum\Zone=\InZoneC \global\TabInZoneCtrue\fi
   \fi
  \global\advance\NextTable \@ne
   \Print{This table has been placed.}
  \else
  \Print{No space available for this table...holding over.}
   \Print{}
   \global\MoreTablesfalse
  \fi
}


\def\CheckFitInZone{%
{\advance\TextSize by -\ItemSIZE
 \advance\TextSize by -\TextLeading
 \ifFirstSingleItem
  \advance\TextSize by \TextLeading
 \fi
 \ifnum\Zone=\InZoneA\relax
  \else \advance\TextSize by -\ZoneBAdjust
 \fi
 \ifdim\TextSize<3\TextLeading \global\ItemFits=\No
 \else \global\ItemFits=\Yes\fi}
}

\def\BeginOpening{%
  \thispagestyle{titlepage}%
  \global\setbox\ItemBOX=\vbox\bgroup%
    \hsize=\PageWidth%
    \hrule height \z@
    \ifsinglecol\vskip 6pt\fi 
}

\let\begintopmatter=\BeginOpening  

\def\EndOpening{%
  \One
  \egroup
  \ifsinglecol
    \box\ItemBOX%
    \vskip\TextLeading plus 2\TextLeading
    \@noafterindent
  \else
    \ItemNUMBER=\z@%
    \ItemTYPE=\Figure
    \ItemSPAN=\Double
    \ItemSTATUS=\InStack
    \JoinStack
  \fi
}


\newif\if@here  \@herefalse

\def\no@float{\global\@heretrue}
\let\nofloat=\relax 

\def\beginfigure{%
  \@ifstar{\global\@dfloattrue \@bfigure}{\global\@dfloatfalse \@bfigure}%
}

\def\@bfigure#1{%
  \par
  \if@dfloat
    \ItemSPAN=\Double
    \TEMPDIMEN=\PageWidth
  \else
    \ItemSPAN=\Single
    \TEMPDIMEN=\ColumnWidth
  \fi
  \ifsinglecol
    \TEMPDIMEN=\PageWidth
  \else
    \ItemSTATUS=\InStack
    \ItemNUMBER=#1%
    \ItemTYPE=\Figure
  \fi
  \bgroup
    \hsize=\TEMPDIMEN
    \global\setbox\ItemBOX=\vbox\bgroup
      \eightpoint\nostb@ls{10pt}%
      \let\caption=\fig@caption
      \ifsinglecol \let\nofloat=\no@float\fi
}

\def\fig@caption#1{%
  \vskip 5.5pt plus 6pt%
  \bgroup 
    \eightpoint\nostb@ls{10pt}%
    \setbox\TEMPBOX=\hbox{#1}%
    \ifdim\wd\TEMPBOX>\TEMPDIMEN
      \noindent \unhbox\TEMPBOX\par
    \else
      \hbox to \hsize{\hfil\unhbox\TEMPBOX\hfil}%
    \fi
  \egroup
}

\def\endfigure{%
  \par\egroup 
  \egroup
  \ifsinglecol
    \if@here \midinsert\global\@herefalse\else \topinsert\fi
      \unvbox\ItemBOX
    \endinsert
  \else
    \JoinStack
    \Print{Processing source for figure \the\ItemNUMBER}%
  \fi
}


\newbox\tab@cap@box
\def\tab@caption#1{\global\setbox\tab@cap@box=\hbox{#1\par}}

\newtoks\tab@txt@toks
\long\def\tab@txt#1{\global\tab@txt@toks={#1}\global\table@txttrue}

\newif\iftable@txt  \table@txtfalse
\newif\if@dfloat    \@dfloatfalse

\def\begintable{%
  \@ifstar{\global\@dfloattrue \@btable}{\global\@dfloatfalse \@btable}%
}

\def\@btable#1{%
  \par
  \if@dfloat
    \ItemSPAN=\Double
    \TEMPDIMEN=\PageWidth
  \else
    \ItemSPAN=\Single
    \TEMPDIMEN=\ColumnWidth
  \fi
  \ifsinglecol
    \TEMPDIMEN=\PageWidth
  \else
    \ItemSTATUS=\InStack
    \ItemNUMBER=#1%
    \ItemTYPE=\Table
  \fi
  \bgroup
    \eightpoint\nostb@ls{10pt}%
    \global\setbox\ItemBOX=\vbox\bgroup
      \let\caption=\tab@caption
      \let\tabletext=\tab@txt
      \ifsinglecol \let\nofloat=\no@float\fi
}

\def\endtable{%
  \par\egroup 
  \egroup
  \setbox\TEMPBOX=\hbox to \TEMPDIMEN{%
    \hss
    \vbox{%
      \hsize=\wd\ItemBOX
      \ifvoid\tab@cap@box
      \else
        \noindent\unhbox\tab@cap@box
        \vskip 5.5pt plus 6pt%
      \fi
      \box\ItemBOX
      \iftable@txt
        \vskip 10pt%
        \eightpoint\nostb@ls{10pt}%
        \noindent\the\tab@txt@toks
        \global\table@txtfalse
      \fi
    }%
    \hss
  }%
  \ifsinglecol
    \if@here \midinsert\global\@herefalse\else \topinsert\fi
      \box\TEMPBOX
    \endinsert
  \else
    \global\setbox\ItemBOX=\box\TEMPBOX
    \JoinStack
    \Print{Processing source for table \the\ItemNUMBER}%
  \fi
}

\def\UnloadZoneA{%
\FirstZoneAtrue
 \Iteration=\z@
  \loop
   \ifnum\Iteration<\LengthOfStack
    \GetItemSTATUS{\Iteration}
    \ifnum\ItemSTATUS=\InZoneA
     \GetItemBOX{\Iteration}
     \ifFirstZoneA \vbox to \BodgeHeight{\vfil}%
     \FirstZoneAfalse\fi
     \unvbox\ItemBOX\ItemSep
     \LeaveStack{\Iteration}
     \else
     \advance\Iteration \@ne
   \fi
 \repeat
}

\def\UnloadZoneC{%
\Iteration=\z@
  \loop
   \ifnum\Iteration<\LengthOfStack
    \GetItemSTATUS{\Iteration}
    \ifnum\ItemSTATUS=\InZoneC
     \GetItemBOX{\Iteration}
     \ItemSep\unvbox\ItemBOX
     \LeaveStack{\Iteration}
     \else
     \advance\Iteration \@ne
   \fi
 \repeat
}


\def\ShowItem#1{
  {\GetItemAll{#1}
  \Print{\the#1:
  {TYPE=\ifnum\ItemTYPE=\Figure Figure\else Table\fi}
  {NUMBER=\the\ItemNUMBER}
  {SPAN=\ifnum\ItemSPAN=\Single Single\else Double\fi}
  {SIZE=\the\ItemSIZE}}}
}

\def\ShowStack{%
 \Print{}
 \Print{LengthOfStack = \the\LengthOfStack}
 \ifnum\LengthOfStack=\z@ \Print{Stack is empty}\fi
 \Iteration=\z@
 \loop
 \ifnum\Iteration<\LengthOfStack
  \ShowItem{\Iteration}
  \advance\Iteration \@ne
 \repeat
}

\def\B#1#2{%
\hbox{\vrule\kern-0.4pt\vbox to #2{%
\hrule width #1\vfill\hrule}\kern-0.4pt\vrule}
}


\newif\ifsinglecol   \singlecolfalse

\def\onecolumn{%
  \global\output={\singlecoloutput}%
  \global\hsize=\PageWidth
  \global\vsize=\PageHeight
  \global\ColumnWidth=\hsize
  \global\TextLeading=12pt
  \global\Leading=12
  \global\singlecoltrue
  \global\let\onecolumn=\relax
  \global\let\footnote=\sing@footnote
  \global\let\vfootnote=\sing@vfootnote
  \ninepoint 
  \message{(Single column)}%
}

\def\singlecoloutput{%
  \shipout\vbox{\PageHead\pagebody\PageFoot}%
  \advancepageno
  \ifplate@page
    \shipout\vbox{%
      \sp@pagetrue
      \def\sp@type{plate}%
      \global\plate@pagefalse
      \PageHead\vbox to \PageHeight{\unvbox\plt@box\vfil}\PageFoot%
    }%
    \message{[plate]}%
    \advancepageno
  \fi
  \ifnum\outputpenalty>-\@MM \else\dosupereject\fi%
}

\def\ItemSep{\vskip\ItemSepamount\relax}

\def\ItemSepbreak{\par\ifdim\lastskip<\ItemSepamount
  \removelastskip\penalty-200\ItemSep\fi%
}


\let\@@endinsert=\endinsert 

\def\endinsert{\egroup 
  \if@mid \dimen@\ht\z@ \advance\dimen@\dp\z@ \advance\dimen@12\p@
    \advance\dimen@\pagetotal \advance\dimen@-\pageshrink
    \ifdim\dimen@>\pagegoal\@midfalse\p@gefalse\fi\fi
  \if@mid \ItemSep\box\z@\ItemSepbreak
  \else\insert\topins{\penalty100 
    \splittopskip\z@skip
    \splitmaxdepth\maxdimen \floatingpenalty\z@
    \ifp@ge \dimen@\dp\z@
    \vbox to\vsize{\unvbox\z@\kern-\dimen@}
    \else \box\z@\nobreak\ItemSep\fi}\fi\endgroup%
}


\def\gobbleone#1{}
\def\gobbletwo#1#2{}
\let\footnote=\gobbletwo 
\let\vfootnote=\gobbleone

\def\sing@footnote#1{\let\@sf\empty 
  \ifhmode\edef\@sf{\spacefactor\the\spacefactor}\/\fi
  \hbox{$^{\hbox{\eightpoint #1}}$}\@sf\sing@vfootnote{#1}%
}

\def\sing@vfootnote#1{\insert\footins\bgroup\eightpoint\b@ls{9pt}%
  \interlinepenalty\interfootnotelinepenalty
  \splittopskip\ht\strutbox 
  \splitmaxdepth\dp\strutbox \floatingpenalty\@MM
  \leftskip\z@skip \rightskip\z@skip \spaceskip\z@skip \xspaceskip\z@skip
  \noindent $^{\scriptstyle\hbox{#1}}$\hskip 4pt%
    \footstrut\futurelet\next\fo@t%
}

\def\footnoterule{\kern-3\p@ \hrule height \z@ \kern 3\p@}

\skip\footins=19.5pt plus 12pt minus 1pt
\count\footins=1000
\dimen\footins=\maxdimen


\def\landscape{%
  \global\TEMPDIMEN=\PageWidth
  \global\PageWidth=\PageHeight
  \global\PageHeight=\TEMPDIMEN
  \global\let\landscape=\relax
  \onecolumn
  \message{(landscape)}%
  \raggedbottom
}


\output{%
  \ifLeftCOL
    \global\setbox\LeftBOX=\vbox to \ZoneBSize{\box255\unvbox\ZoneBBOX}%
    \global\LeftCOLfalse
    \MakeRightCol
  \else
    \setbox\RightBOX=\vbox to \ZoneBSize{\box255\unvbox\ZoneBBOX}%
    \setbox\MidBOX=\hbox{\box\LeftBOX\hskip\ColumnGap\box\RightBOX}%
    \setbox\PageBOX=\vbox to \PageHeight{%
      \UnloadZoneA\box\MidBOX\UnloadZoneC}%
    \shipout\vbox{\PageHead\box\PageBOX\PageFoot}%
    \advancepageno
    \ifplate@page
      \shipout\vbox{%
        \sp@pagetrue
        \def\sp@type{plate}%
        \global\plate@pagefalse
        \PageHead\vbox to \PageHeight{\unvbox\plt@box\vfil}\PageFoot%
      }%
      \message{[plate]}%
      \advancepageno
    \fi
    \global\LeftCOLtrue
    \CleanStack
    \MakePage
  \fi
}


\Warn{\start@mess}


\catcode `\@=12 



\def\hp{H_{\rm P}}
\def\chid{\chi_{\rm D}}
\def\deldel{(\nabla-\nabla_{\rm ad})}
\def\deldeln{\nabla-\nabla_{\rm ad}}
\def\delmu{\nabla_\mu}
\def\wtb{\overline{w\theta}}
\def\msun{M_{\sun}}
\def\al{\alpha_{\rm L}}
\def\tauh{\tau_{\rm H}}
\def\tauhe{\tau_{\rm He}}
\def\mh{M_{\rm H}}
\def\mhe{M_{\rm He}}

\onecolumn
\Autonumber

\begintopmatter
\title{Double-Diffusive Mixing-Length Theory, Semiconvection,
and Massive Star Evolution}
\author{Scott~A.~Grossman and Ronald E. Taam}
\affiliation{Northwestern University, Dearborn Observatory, 2131 Sheridan Rd.,
Evanston, IL 60208}

\shortauthor{Grossman and Taam}
\shorttitle{Double-Diffusive Mixing-Length Theory}

\abstract{Double-diffusive convection refers to mixing where the
effects of thermal and composition gradients compete to determine the
stability of a fluid.  In addition to the familiar fast convective
instability, such fluids exhibit the slow, direct salt finger
instability and the slow, overstable semiconvective instability.
Previous approaches to this subject usually have been based on linear
stability analyses.  We develop here the nonlinear mixing-length
theory (MLT) of double-diffusive convection, in analogy to the more
familiar MLT for a fluid of homogeneous composition.  We present
approximate solutions for the mixing rate in the various regimes, and
show that the familiar Schwarzschild and Ledoux stability criteria are
good approximations to the precise criteria in stellar interiors.

We have implemented the self-consistent computation of the temperature
gradient and turbulent mixing rate in a stellar evolution code and
solved a diffusion equation to mix composition at the appropriate
rate.  We have evolved $15\msun$ and $30\msun$ stars from the zero-age
main sequence to the end of core He-burning.  Semiconvective mixing is
fast enough to alter stellar composition profiles on relevant time
scales, but not so fast that instantaneous readjustment is
appropriate.}

\keywords{convection--hydrodynamics--instabilities--stars: evolution and
interiors}

\maketitle

\section{Introduction}

Double-diffusive instabilities arise when the transport of two
different properties compete against each other to dominate the
stability of a fluid.  In stars, heat and composition are the two
quantities transported by mixing.  Semiconvection occurs when a
thermally driven fluid has a composition gradient that opposes the
instability.  This subject also goes by the names of thermosolutal and
thermohaline convection, as motivated by its relevance to
oceanography, where varying salinity causes composition gradients.  In
stars, nuclear burning provides the source of heat for thermal
driving.  It also causes composition changes as elements are
transformed into species of higher molecular weight, leading to the
possibility of a stabilizing composition gradient.

A fluid of homogeneous composition is unstable to convection if its
temperature gradient is steeper than the adiabatic gradient,
$$\deldeln>0,\eqno(1)$$ where $\nabla=\partial\ln T/\partial\ln P$ is
the logarithmic temperature gradient and $\nabla_{\rm ad}$ gives the
temperature change of an adiabatic displacement. This is the famous
Schwarzschild criterion.  In stars, convection is very rapid compared
to an evolutionary time scale, and small composition gradients mix to
homogeneity virtually instantaneously.

If a star is Schwarzschild unstable in a region with a significant
composition gradient (such as left by the retreating convective core
of a main sequence star), what happens?  If the composition gradient
that opposes the thermal instability satisfies the relation
$$\deldel-\delmu<0,\eqno(2)$$ the rapid growth of convective motions
is stabilized; this does not preclude instabilities of slower growth.
$\delmu=\partial\ln\mu/\partial\ln P$ is the nondimensional
composition gradient, where the composition is measured by the
molecular weight $\mu$.  This is the Ledoux stability criterion
(Ledoux 1947).  If the fluid is Ledoux stable, but Schwarzschild
unstable, can the composition gradient be sustained indefinitely?  It
cannot, but the mixing is much slower than the fast convective mixing
of a convective zone.

Consider a fluid blob displaced upward from its equilibrium position
in a background with temperature and composition gradients such that
the fluid is Ledoux stable and Schwarzschild unstable.  The displaced
blob will be hotter than its environment because of the Schwarzschild
instability, and thermal buoyancy will drive it upward.  It will also
be heavier than its environment, and it will feel a force downward.
Because the fluid is Ledoux stable, the downward force wins and there
is a net restoring force, so that the blob oscillates around its
equilibrium position.  If there is nonzero rate of thermal diffusion,
then while the blob is on its upward excursion where it is hotter than
its environment, it loses heat and returns to the equilibrium position
with a lower temperature than it began with.  When it then makes its
downward excursion, negative buoyancy will make it travel somewhat
deeper than it would have otherwise.  While on the downward excursion,
it gets hotter and returns to the equilibrium position somewhat too
hot and has an excess buoyancy.  Thus, the amplitude of the
oscillation grows.  The phase lag between the temperature and velocity
oscillations is responsible for the work that increases the amplitude
of oscillation.  This vibrational instability or overstability grows
on the time scale of thermal diffusion.  It is what those who study
the stability of fluids call semiconvection (Kato 1966; Baines \& Gill
1969; Grossman, Narayan, \& Arnett 1993, hereafter GNA), and is the
description that can be found in text books (Kippenhahn \& Weigert
1991) and review articles (Spiegel 1972).

Semiconvection is known to occur in massive stars and horizontal
branch stars.  The idea of semiconvection in stars originated in a
classic paper by Schwarzschild \& H\"arm (1958; cf. also Sakashita \&
Hayashi 1959).  They noticed that in massive main sequence stars
($M\ga 10M_\odot$) where electron scattering is the dominant source of
opacity, the opacity is larger in the H rich envelope outside the He
enriched convective core.  Thus, outside the Schwarzschild boundary
(as defined by the fully mixed core), the H rich envelope is also
Schwarzschild unstable.  The core, however, was growing, bringing the
envelope back to stability.  The issue of whether the envelope was
stable or unstable was resolved by establishing a zone of partial
mixing, where He rich material was mixed into the envelope until it
was neutrally stabile.\footnote{*}{Although Schwarzschild
\& H\"arm thought the H-burning convective core grows during main
sequence evolution, causing a composition jump at the core boundary,
calculations since have shown that the convective core retreats
(Simpson 1971), leaving behind a composition gradient.  A
redistribution of composition until neutral stability is reached
occurs in either case.}  Whether the condition for neutral stability
is expressed by the Schwarzschild or Ledoux criterion was a point of
contention about 30 years ago (e.g., Spiegel 1969; Gabriel 1969), but
the former is now accepted.  This redistribution of composition is
what most authors in stellar evolution refer to as semiconvection,
irrespective of the specific nature of the instability.  The
importance of semiconvection to massive star evolution was manifest
with SN 1987A, which exploded as a blue supergiant, rather than as a
red supergiant as expected.  Langer, El Eid, \& Baraffe (1989) and
Arnett (1991) found that whether a massive star explodes while it is
blue or while it is red depends sensitively on the semiconvective rate
of mixing.

What is the evidence for formation of semiconvective zones (SCZs),
regions of partial mixing where the envelope is Schwarzschild
unstable?  Comparisons of observations of the numbers of blue versus
red supergiants with theoretical evolutionary calculations of massive
stars favor instantaneous mixing out to the Ledoux boundary only,
i.e. the boundary of the retreating H-burning core.  Complete mixing
out to the Schwarzschild boundary where the H enriched envelope
eventually becomes stable is disfavored (Stothers \& Chin 1992a,b,
Stothers \& Chin 1994).  A model of semiconvective mixing suggests
that the mixing is slow enough that instantaneous mixing does not
happen in the SCZ and that the core mass is significantly smaller than
in models with instantaneous mixing to the Schwarzschild boundary
(Langer, El Eid, \& Fricke 1985; Langer 1991).

GNA used a Boltzmann transport description of convection to derive the
known results of mixing-length theory (MLT) (B\"ohm-Vitense 1958).
Their results suffer from the same lack of rigor as all formulations
of MLT, namely that damping from the turbulent cascade is approximated
by damping on the largest scale only, the mixing length, which is
itself unknown.  Their method has the advantage that standard MLT can
be extended to more complicated promblems easily.  They demonstrated
that their method reproduces the double-diffusive linear
instabilities, but the nonlinear MLT of double-diffusive convection
was not developed fully.  We extend the results of GNA on
double-diffusive instabilities here to formulate the local MLT of
double-diffusive convection, that is, the nonlinear point where the
linear instabilities saturate.  Our results apply in all stability
regimes--convective, semiconvective, and salt finger.\footnote{\dag}
{The salt finger instability occurs when a fluid is stable by the
Ledoux criterion, but driven by the composition gradient instead of by
the temperature gradient.} Some work has been done previously that
gives similar results for various subsets of this problem.  Linear
stability analyses of semiconvection have been discussed by Kato
(1966), Langer, Sugimoto, \& Fricke (1983), Eggleton (1983), and
Nakakita \& Umezu (1994).  GNA followed an approach similar to Xiong
(1981) and Eggleton (1983) by writing the hydrodynamic equations for
correlations of perturbations.  The nonlinear MLT with composition
dependence was developed by Umezu \& Nakakita (1988) and Umezu (1989),
but there was some confusion regarding the relevant solutions.  The
salt finger instability has been discussed by Ulrich (1971) and
Kippenhahn, Ruschenplatt, \& Thomas (1980).  Relevant nonlocal studies
have been made by Shibahashi
\& Osaki (1976) and Gabriel \& Noels (1976), who computed the
oscillatory modes of massive stars, and by Gough \& Toomre (1982), who
performed a modal analysis of semiconvection.  We do not develop the
nonlocal MLT theory here.

It has been most common to treat semiconvection in stellar evolution
codes using an iterative scheme that maintains convective neutrality
instantaneously.  We have implemented the double-diffusive local MLT
in a stellar evolution code to test the validity of instantaneous
mixing to neutrality.  To our knowledge, only Langer, El Eid, \&
Fricke (1985) and Langer (1991) have done something comparable by
solving a diffusion equation for composition evolution using the
semiconvective diffusion rate of Langer (1983).  Langer's convective
time scale is derived from a linear stability analysis of the growth
of semiconvective modes, whereas our time scale depends on the
nonlinear velocity where the growth saturates.

We note that in the laboratory, semiconvective mixing evolves such
that regions of rapid mixing are separated by boundary layers, across
which the composition varies in discrete steps.  The number of layers
grows on a thermal diffusion time scale, until the fluid is thoroughly
mixed.  This is a consequence of a nonlinear instability whose onset
is at smaller $\deldeln$ than for the linear instability (Proctor
1981).  This process has been modeled by Spruit (1992).  Our methods
cannot reproduce the composition steps and boundary layers of
laboratory experiments, nor could any mixing length approximation to
stellar convection.  We think it likely that semiconvection in stars
may be more turbulent and less ordered than the laboratory experiments
due to the extremely high Reynolds numbers and low Prandtl numbers of
stellar fluids (Gabriel 1970; Stevenson 1977, 1979).

In \S2, we derive the MLT of fluids with composition gradients.  We
review results of the linear stability analysis and discuss the
nonlinear point where these instabilities saturate.  We show how to
use flux conservation to obtain self-consistent temperature gradients
and turbulent velocities simultaneously.  In \S3, we describe the
implementation of the physics into the stellar evolution code
described by Eggleton (1971, 1972).  We discuss examples of massive
star evolution using the extended MLT and compare results to those
from the unmodified code and from other authors.  In \S4 we summarize
and discuss the results and the possibility of using the extended MLT
for horizontal branch evolution.

\section{The Extended MLT}

In this section we review the essential results of the linear
stability analysis discussed in greater detail by GNA.  We examine the
nonlinear solutions for the turbulent velocity at which the linear
growth saturates and show how to compute the temperature gradient
$\deldeln$ and turbulent velocity simultaneously.

\subsection{The Local Stability Diagram}

\beginfigure{1}\vskip10cm
\caption{{\bf Figure 1.} The $\deldeln$ vs. $\delmu$ stability plane,
with the various stability regimes labeled.  The dotted curve
separates solutions to eq. 3 that are purely real from ones that admit
complex roots.  It is the analog of the Ledoux criterion.  Note,
however, that no combination of parameters could reproduce the
classical picture precisely, where the semiconvective regime is
bounded by the Ledoux and Schwarzschild lines.  In the semiconvective
regime, the dominant mode of growth is overstable.  In the salt finger
regime, the dominant mode of growth is direct, even though oscillatory
modes are present also.  We use parameters taken from just outside the
H-burning core of a $30\msun$ star.}
\endfigure

As shown by GNA (Appendix C), in a fluid characterized by a
superadiabatic gradient $\deldeln$ and a composition gradient
$\delmu$, if linear perturbations vary in time like $e^{st}$, then the
eigenvalues $s$ of linear growth satisfy the cubic equation
$$s^3+as^2+bs+c=0.\eqno(3)$$ The coefficients are given by
$$a=A+D+F,\eqno(4a)$$
$$b=(AD+AF+DF)-{g\alpha\over\hp}\deldel+{g\phi\over\hp}\delmu,\eqno(4b)$$
$$c=-{Fg\alpha\over\hp}\deldel+{Dg\phi\over\hp}\delmu+ADF,\eqno(4c)$$
where $g$ is gravity, $\hp$ is a pressure scale height, and
$\alpha=-(\partial\ln\rho/\partial\ln T)_{P,\mu}$ and
$\phi=(\partial\ln\rho/\partial\ln\mu)_{P,T}$ are constants derived
from the equation of state.  The parameters $A$, $D$, and $F$ define
the diffusion rates of viscosity, heat, and composition according to
$$A=10\nu_{\rm mic}/3\ell^2,\eqno(5a)$$
$$D=3\chi/\ell^2,\eqno(5b)$$
$$F=3\chid/\ell^2,\eqno(5c)$$ where $\nu_{\rm mic}$, $\chi$, and
$\chid$ are measured in $\rm cm^2/s$. Although GNA allowed for several
mixing lengths for the various turbulent diffusion rates and for
potentially different horizontal and vertical dimensions of a
convective eddy, here we take all mixing lengths equal to the single
value $\ell$.  (Clearly it would be more appropriate for the salt
finger instability to worry about the different horizontal and
vertical dimensions of an eddy using the original definitions of GNA.)
Results essentially equivalent to equation (3) have been derived by
many authors previously, beginning with Kato (1966) and Baines \& Gill
(1969).

The detailed criteria for instability depend on the diffusion rates of
heat, composition, and momentum.  In stars, the exchange of heat by
radiative diffusion is always faster than the diffusion of
composition, and the diffusion of momentum by molecular viscosity is
negligible.  As shown by GNA (see also Kato 1966, Baines \& Gill
1969), the $\deldeln$ versus $\delmu$ parameter space is divided by
lines of critical stability, across which the roots of equation (3)
change their character.  If the roots are all real, the stability
changes when the largest root equals zero.  This transition is defined
by
$$Fg\alpha\deldel-Dg\phi\delmu=\hp ADF.\eqno(6)$$
Another possibility is that the dominant root is a complex conjugate
pair.  The stability changes when the real part of this root equals zero.
This transition is defined by
$$(A+D)g\alpha\deldel-(A+F)g\phi\delmu=\hp(A+D)(A+F)(D+F).\eqno(7)$$
The curve that separates solutions of these two types, i.e. pure
exponential growth/decay of linear perturbations from oscillatory
growth/decay is given by
$$\left({b\over 3}-{a^2\over 9}\right)^3+\left({ab-3c\over 6}\right)^2=0.
\eqno(8)$$
A more complete discussion of stability behavior can be found in GNA.
In the limit that the diffusion rate $D$ goes to zero and $A\ll D$,
$F\ll D$, equations (6), (7), and (8) simplify to the Rayleigh-Taylor,
Schwarzschild, and Ledoux criteria, respectively.  The regimes of
semiconvective and salt finger instability exist only as a consequence
of a nonzero rate of thermal diffusion.

In Figure 1, the critical lines defined by equations (6) and (7) are
drawn as solid lines, and the curve defined by equation (8) is dotted.
These lines correspond very nearly to the stability criteria
$\deldeln=0$ and $\delmu=0$ if the thermal diffusion is much greater
than the composition diffusion ($D\gg F$) and viscosity (A) is
negligible, as is true in stellar interiors.  GNA showed that the
region of overstability is bounded by the dotted line and the line of
equation (7), the line that approximates the Schwarzschild criterion.
The dotted curve has a cusp near the origin, and does not appear to
resemble the Ledoux criterion, which would trace a diagonal line from
upper left to lower right in this figure.  Indeed, for no combination
of parameters is the fluid stable by the Ledoux line, but unstable by
the Schwarzschild line.  This lead GNA to suggest that the text book
description of semiconvection may not be appropriate for use in
realistic calculations.  We demonstrate below that this conclusion was
incorrect.

\subsection{The Nonlinear Turbulent Velocity}

We begin by writing the equation for the turbulent velocity $\sigma$,
$$\sigma^2\Big[(A+D+2B\sigma)g\alpha\deldel
-(A+F+2B\sigma)g\phi\delmu
-\hp(A+D+2B\sigma)(A+F+2B\sigma)(D+F+2B\sigma)\Big]$$
$$\times\Big[(F+B\sigma)g\alpha\deldel-(D+B\sigma)g\phi\delmu
-\hp(A+B\sigma)(D+B\sigma)(F+B\sigma)\Big]=0,\eqno(9)$$ (GNA,
eq. 7.34).\footnote{*}{Eq. (3) is cubic because it derives from the
three equations of motion for the perturbations of velocity, $w$
temperature, $\theta$, and composition, $\nu$.  Eq. (9), ignoring the
leading $\sigma^2$, is sixth order because steady convection is
described by the six equations for the correlations $\overline{w^2}$,
$\overline{w\theta}$, $\overline{\theta^2}$, $\overline{w\nu}$,
$\overline{\theta\nu}$, and $\overline{\nu^2}$.  The leading term
$\sigma^2=\overline{w^2}$ admits the trivial solution of the equations,
where all correlations are zero.} We define
$$B=2/\ell,\eqno(10)$$ so that $B\sigma$ is the turbulent damping
rate, in analogy to the microscopic rates given by $A$, $D$, and $F$.
Physically meaningful roots to this equation are real and
non-negative.  Not all these roots will be stable equilibria of the
time dependent moment equations of GNA from which equation (9) was
derived.  In general, the fluid will seek out the most turbulent
equilibrium state, and only this solution will be stable.  The leading
factor $\sigma^2$ shows that $\sigma=0$ is always an equilibrium
solution for a fluid; indeed, it is the only physical solution if the
fluid is in the stable regime.  The rest of equation (9) is the
product of two terms, both cubic in $\sigma$.  Each term will have at
least one real solution.  If one or more are positive, the equilibrium
state will evolve to the largest positive solution, and the solution
$\sigma=0$ will be an unstable equilibrium (cf. the stability analyses
of GNA and Nakakita \& Umezu 1994).

We consider approximate solutions to equation (9) in the convective,
salt finger, and semiconvective regimes.  In the regime of convective
instability, all roots are real (see below).  The largest positive
root comes from the second term in brackets in equation (9), which can
be expanded to read
$$(B\sigma)^3+a(B\sigma)^2+b(B\sigma)+c=0.\eqno(11)$$ Thus, in the
convective regime defined by equation (8), all roots $\sigma$ are
real, just as for the eigenvalues $s$.\footnote{\dag}{One way to
estimate $\sigma$ from the linear analysis is to replace $s$ in
eq. (3) by $\sigma/\ell\sim B\sigma$ (Eggleton 1983).  In the regime
of overstability where $s$ is complex, $\sigma$ will be too.  Eggleton
(1983) suggested one should use the real part of this complex $\sigma$
to describe the mixing rate.  As we show, this supposition is correct,
although in the sixth order eq. (9), the appropriate solution $\sigma$
is strictly real, as must be any value of the rms turbulent velocity.}
Equation (11) also illustrates the close relationship between the
linear stability analysis and the nonlinear MLT.  This is not
surprising since the MLT is derived by assuming eddy-damping rates
balance growth rates given by $s$ in the linear approximation.

To approximate the root $\sigma$ in the convective regime where all
roots are real, we simplify the calculation by setting the rate of
molecular composition diffusion, $F=0$, and consider the limit of
efficient convection where the turbulent transport of heat dominates
radiative transport, $B\sigma/D\gg 1$.  In this limit, $c\ll
bB\sigma$, and equation (11) can be reduced to a quadratic relation,
with positive real solution
$$\sigma\approx\left[{g\alpha\deldel-g\phi\delmu\over{\hp
B^2}}\right]^{1/2}.\eqno(12)$$ If $\delmu=0$, this approximation is
precisely the local estimate of the turbulent velocity in the
efficient convective regime as derived in GNA.  This shows that the
rate of turbulent diffusion $\ell\sigma\propto\ell^2$. In the general
case that $\delmu$ is not equal to zero, it is evident that across the
Ledoux line, equation (12) becomes complex and does not give a
physical solution for the semiconvective regime.  Numerical solutions
of the cubic equation (11) show that across the curve defined by
equation (8), the two nontrivial real roots merge into a complex
conjugate pair with a small real part.\footnote{*}{Across the Ledoux
curve, the second cubic factor has roots in the pattern $2R_1$,
$2R_2\pm i2I_2$, and the first cubic factor has roots in the pattern
$2R_2$, $(R_1+R_2)\pm iI_2$.  Thus, the real part of the complex
conjugate pair approximated by eq. (12) as $\pm i2I_2$ is the same as
the real root of the first cubic factor.  In the salt finger regime,
$2R_1>0$ and $2R_2<0$.  In the semiconvective regime, $2R_1<0$, and
the physical root is $2R_2>0$.}

Crossing the Ledoux curve in the direction of the salt finger
instability, the small real root of equation (11) is the largest real
root of the higher order equation (9).  Factoring out the complex
conjugate pair, one can show that the value of the real root
$$\sigma\approx -\left({D\over B}\right)
{\phi\delmu\over{\phi\delmu-\alpha\deldel}}.\eqno(13)$$ In the salt
finger regime where the molecular weight profile is inverted,
$\delmu<0$ and $\sigma$ is positive.  This is the result from the
linear stability analysis of Kippenhahn et al. (1980).

Crossing the Ledoux curve in the direction of the semiconvective
instability where $\delmu>0$, clearly equation (13) does not give a
physical solution.  In this case, the physical solution comes from the
first cubic factor of equation (9).  This real root equals the real
part of the complex conjugate pair.  Equation (12) approximates the
imaginary part of this pair.  If we take the limit of $B\sigma/D\gg 1$
in the first factor of equation (9), then the quadratic and zeroth
order terms are small, with the cubic and linear terms dominating.  In
this case, the real root is small, and the dominant roots are a
complex conjugate pair.  Clearly the solution we require is the small
real root.  Factoring out the approximate complex roots, we can show
that the real root must be
$$\sigma\approx\left({D\over{2B}}\right)
{\alpha\deldel\over{\phi\delmu-\alpha\deldel}}.\eqno(14)$$ Numerical
solutions show this approximation is good, even if $B\sigma/D\ll 1$,
so long as $\deldeln$ is not close to the critical line of equation
(6), i.e., $\deldel/\deldel_{\rm crit}\gg 1$, where $\deldel_{\rm
crit}= F\phi\delmu/D\alpha$.  Taking the semiconvective diffusion rate
as $\ell\sigma$, equation (14) essentially gives Langer et al.'s
(1983) result, and illustrates the essential feature that the rate of
semiconvective mixing depends on the thermal diffusion rate $D$.
Langer et al. (1983), however, multiplied their diffusion rate by an
arbitrary constant, $\al$, to account for uncertainties in mixing
length theory and the translation of a linear growth rate to a rate of
nonlinear mixing.  The effect of semiconvection is to mix regions
toward neutral stability, so that $\deldeln$ approaches zero.

Equation (14) shows $\sigma\propto\ell^{-1}$, so that the diffusion
rate, $\ell\sigma$, does not depend on the value of the mixing length.
Thus, for decreasing $\ell$, $\sigma$ increases, approaching the
solution approximated by equation (12).  It reaches a maximum, and for
$\ell$ yet smaller, $\sigma$ decreases.  As long as a fluid is in the
semiconvective regime, the rate of turbulent mixing is independent of
the particular choice of mixing length.  For sufficiently small
$\ell$, the fluid is actually in the convective regime, but the
diffusion rate is even smaller.

The complex condition, equation (8), will approximate the Ledoux
criterion in the limit that the term $b/3$ dominates the others, that
is, in the limit that
$${g\alpha\deldel\over\hp}-{g\phi\delmu\over\hp}\gg {\cal O}(A,D,F)
{\cal O}(A,D,F),\eqno(15)$$ where the right-hand-side represents all
permutations of the microscopic diffusion rates for momentum, heat,
and composition.  If a convective zone, bordering on a semiconvective
region, is in the regime of efficient convection, we can use equation
(12) to show that the left-hand-side of equation (15) $\sim
B^2\sigma^2$ according to standard MLT.  But
$B^2\sigma^2\sim\sigma^2/\ell^2\sim 1/\tau_{\rm conv}^2,$ where
$\tau_{\rm conv}$ is the the characteristic time for the turnover of a
convective eddy.  On the other hand, the right-hand-side of equation
(15) can be represented as $1/\tau_{\rm mic}^2$, where $\tau_{\rm
mic}$ is the characteristic damping time by microscopic diffusion
processes.  Thus, equation (15) translates into the condition
$$\tau_{\rm conv}\ll\tau_{\rm mic}.\eqno(16)$$ Thus, when the
convection zone is in the efficient regime where fluid blobs move
nearly adiabatically and losses by microscopic diffusion are
negligible, the transition to semiconvection occurs across a boundary
that closely approximates the Ledoux criterion.  Only when a SCZ
borders on a region of inefficient convection is it important to worry
about the detailed cubic relation of equation (8).  This never happens
in stellar interiors, and thus the Ledoux criterion is an excellent
approximation to the exact result.  It is not necessary to compute the
detailed stability criteria from equations (6)--(8), as GNA
erroneously claimed.

The stability diagram is shown again in Figure 2, but extended to
values of $\deldeln$ and $\delmu$ relevant for the convective and
semiconvective zones of stars.  Contours of constant turbulent
velocity $\sigma$ are drawn.  In the convective core of a massive star
(see \S3), typical convective velocities are of order $10^5\,\rm
cm/s$.  For a pressure scale height or mixing length of order
several$\times 10^{10}\rm cm$, a characteristic mixing time is
$\tau_{\rm conv}\sim 0.01\,\rm yr$.  Since the size of the convective
core is only a few times larger than the pressure scale height,
instantaneous mixing is an excellent approximation on evolutionary
time scales.

Crossing into the semiconvective zone, the mixing rate falls
dramatically.  In the SCZ of a massive star, typical velocities are of
order $10^{-3}\,\rm cm/s$, giving a mixing time $\tau_{\rm conv}\sim
10^6{\rm yr}$.  The size of the SCZ is of order a pressure scale
height, so the time scale to mix the SCZ is only a few times shorter
than the evolutionary time scale.  In fact, since only a partial
redistribution of composition is required to suppress the mixing, the
time scale to mix to neutrality is at least a factor of several
shorter.  Nevertheless, it is clear that instantaneous mixing is not
appropriate in the semiconvective region.

\subsection{Flux Conservation}

\beginfigure{2}\vskip12.6cm
\caption{{\bf Figure 2.}  The stability diagram, just as in
Fig. 1, except extended out by many orders of magnitude to values of
$\deldeln$ and $\delmu$ relevant for stars.  On this scale, the dotted
line resembles the classical Ledoux criterion to excellent
approximation.  This line is not at a $45^\circ$ angle because the
coefficient of thermal expansion $\alpha\approx 1.8$ due to radiation
pressure.  The light, solid curves are contours of constant turbulent
velocity $\sigma$.  Contour levels are $10^{-2}$, $10^{-1}$, $10^0$,
$10^7$, $2\times 10^7\,\rm cm/s$.  The rate of turbulent mixing
changes very rapidly across the Ledoux line and is much slower in the
semiconvective and salt finger regimes.  The open symbols correspond
to examples of the hypothetical radiative temperature gradient of
$\nabla_{\rm Rad}-\nabla_{\rm ad}=0.05$ and $\delmu=0,\,0.025,\,0.15$.
The corresponding solid symbols show the solutions to eq. 17 for the
true temperature gradient $\deldeln$.  In the SCZ, negligible flux is
carried by turbulence, and the temperature gradient is very nearly
equal to the radiative gradient.  This extremely small difference has
been exaggerated here.}
\endfigure

If convection carries some fraction of the energy flux, the true
temperature gradient $\nabla$ is modified from the value $\nabla_{\rm
Rad}$ it would have in the absence of convection.  The true
temperature gradient can be found by writing the equation for flux
conservation,
$$\nabla_{\rm Rad}-\nabla_{\rm ad}=\deldel+\hp\wtb/T\chi.\eqno(17)$$
The term $\wtb$ is the correlation between turbulent velocity and
turbulent temperature excess, and measures the convective flux.  It
can be written in terms of the turbulent velocity $\sigma$.  By
solving the local moment equations (GNA, eqs. 7.28, 7.31--33), we
obtain

$$\wtb={ \left[\alpha\deldel+{\hp\over g}(D+F+2B\sigma)
\left(A+F+2B\sigma+{g\phi\over\hp}{\delmu\over{(F+B\sigma)}}\right)\right]
{(A+B\sigma)\over{(D+F+2B\sigma)}}-\phi\delmu \over
{ \alpha\deldel+{\hp\over g}(D+F+2B\sigma)
\left(A+F+2B\sigma+{g\phi\over\hp}{\delmu\over{(F+B\sigma)}}\right)
-\phi\delmu }} {T\over{\alpha g}}(D+F+2B\sigma)\sigma^2.\eqno(18)$$ In
the limit that $F=\delmu=0$, one can show that equations (17) and (18)
simplify to the standard mixing length result for a homogeneous fluid
(cf.  GNA eqs. 6.16 and 6.19).  For a given value of $\delmu$, only
one value of $\deldeln$ gives $\sigma$ and $\wtb$ (by eqs. 9 and 18)
that satisfies flux conservation (eq. 17).

In Figure 2 we show a few solutions to equation (17).  We use
parameters appropriate to the SCZ of a $30\msun$ main sequence star,
but choose the total flux $\nabla_{\rm Rad}$ and composition gradient
$\delmu$ ourselves for the sake of example.  We choose a total flux
$\nabla_{\rm Rad}-\nabla_{\rm ad}=0.05$ and composition gradients
$\delmu=0,\,0.025,\,0.15$.  For $\delmu=0$, the fluid is homogeneous
and convectively unstable.  Convection is efficient and carries nearly
the total energy flux.  The temperature gradient is very nearly
adiabatic, $\deldeln=3.3\times 10^{-7}$, and $\sigma=3.5\times
10^4\,\rm cm/s$.  For $\delmu=0.025$, the fluid is again convectively
unstable and $\deldeln=0.014\approx\phi\delmu/\alpha$, almost on the
Ledoux line.  The temperature gradient is as nearly adiabatic as it
can be without the convective efficiency dropping precipitously.  In
stars, however, this situation does not occur since mixing is fast and
is thought to evolve to the $\delmu=0$ configuration nearly
instantaneously.  Finally, for $\delmu=0.15$, the fluid is in the
semiconvective regime, where $\deldeln\approx\nabla_{\rm Rad}
-\nabla_{\rm ad}$ to high precision (the difference is exaggerated in
Fig. 2) and $\sigma=0.13\,\rm cm/s$.  In this case, turbulent mixing
is much slower, and convection carries a negligible fraction of the
flux.  We note that in the salt finger regime, the turbulent heat flux
is actually reversed from the convective and semiconvective regimes,
but is negligibly small, like in the semiconvective region.  In this
case, the temperature gradient is also very nearly radiative.

\section{Tests of the Extended MLT}

To demonstrate the feasibility of using the extended MLT in stellar
evolution calculations and to understand its consequences, we have
implemented it in the stellar evolution code of Eggleton (1971, 1972,
1973).  In addition to replacing the standard MLT (i.e., for
homogeneous composition) with a new routine to self-consistently
compute the temperature gradient and mixing rate, we have added a
routine to mix the composition at the appropriate rate.  The
modifications are described below.  We evolve $15\msun$ and $30\msun$
stars, characterized by an initial composition $X=0.7$, $Z=0.02$, from
the zero-age main sequence to the end of core He-burning.  The
constant $\alpha$ is computed from the equation of state, and can be
somewhat larger than unity, mainly because of radiation pressure.  The
constant $\phi=1$.  The rate of thermal diffusion $D$ is computed from
the expression for radiative diffusion, and viscosity and composition
diffusion are zero, $A=F=0$.

Our study is primarily a differential investigation of the
consequences of using the new mixing scheme.  Detailed comparisons
with other authors must be made with caution, since various other
physics may not be the same.  Nevertheless, we compare our results
with those in Langer, El Eid, \& Fricke (1985), who also evolved
$15\msun$ and $30\msun$ stars for the similar composition $X=0.701$,
$Z=0.019$.  They used a semiconvective diffusion rate that essentially
is given by the approximation of equation (14), but multiplied by a
coefficient, $\al$ that allowed them to vary the rate of
semiconvective mixing by orders of magnitude.

\subsection{Code Modification}

In its original form, the Eggleton code identifies a region outside
the H-burning convective core of sufficiently massive stars as
unstable according to the Schwarzschild criterion, but stabilized by
the composition gradient created by the retreating convective core.
Using a somewhat contrived mixing rate, this semiconvective region is
slowly mixed.  Indeed, as we illustrate below, the mixing in this zone
is so slow that composition readjustment does not occur on
evolutionary time scales, and at no time during main sequence
evolution does this region reach neutral stability.  Furthermore, the
mixing rate does not even depend on the composition gradient and does
not contain the physics of semiconvection (Eggleton 1972).

We modified Eggleton's code to use the extended local MLT to compute
the temperature gradient and convective mixing rate in semiconvective
and convective regions.  To mix the composition, the original code
solves a diffusion equation for composition evolution.  The diffusion
time, however, is not derived from the convective time scale.  In
fact, the code is known not to converge for mixing times that are too
short (Eggleton 1972), and it is not possible to use our newly
computed mixing times here.  The minimum mixing time that can be used
is large enough (around $10^5\,\rm yr$) that convective regions
clearly are not mixed instantaneously.  The consequences of this can
be seen in some short-lived phases of evolution that turn out to be of
crucial importance for evolution.  In particular, post-main sequence
evolution depends sensitively on the treatment of mixing.

To remedy this inability to modify the mixing time in the code as
appropriate, we have added a diffusion routine that mixes composition
after each evolutionary step.  The routine is implicit, and accuracy
is best if the diffusion equation is solved on a grid of constant
$\Delta r$.  This requires that the diffusion grid have more points
than the evolution code, and we interpolate back and forth between the
two as required.  During main sequence evolution, the diffusion grid
needs only about 15 times more points than the stellar evolution code,
for which we used 200 mesh points.  Later in the giant phases, the
diffusion grid may require 400 times more points.

The new diffusion routine is decoupled from the rest of the evolution
equations, and this can lead to convergence difficulties.  Therefore,
to maintain numerical stability, we continue to use also the coupled
diffusion as in the original code.  In both convective and
semiconvective regions, the mixing times in the decoupled diffusion
routine are generally shorter than in the coupled equation, so the
mixing time is determined by the decoupled routine, with the original
diffusion only providing numerical stability.

\subsection{$30\msun$ Evolution}

\beginfigure{3}\vskip20.6cm
\caption{{\bf Figure 3.}  The convective structure of the $30\msun$
model for three schemes of convective mixing.  Case A uses only the
standard mixing of Eggleton's code, but the temperature gradient
derived from the extended MLT.  This result is virtually identical to
that from the unmodified Eggleton code.  Case B includes mixing at the
rate derived from the extended MLT.  There is nearly instantaneous
mixing of convective regions, and mixing to convective neutrality in
semiconvective regions.  Case C uses the standard MLT temperature
gradient and mixing rate derived by setting the composition gradient
to zero in the MLT routine.  This is equivalent to using only the
Schwarzschild criterion for convection.  Convective regions are
contoured by a dotted line (which, unfortunately, looks solid in some
places, particularly the convective envelope).  Semiconvective regions
are contoured by a solid line.  In Case A, this is a region of growing
size outside the H-burning core, and a small region near the start of
He core burning (at about $20-25\msun$ and age of $5.6\times 10^6\rm
yr$).  In Case B, semiconvection occurs in the broken-up cells outside
the H-burning core.  In both these cases, there is semiconvection in a
thin shell surrounding the convective core.  In this and all
subsequent figures, we use a mixing length $\ell=1.5\hp$.}
\endfigure

\beginfigure{4}\vskip20cm
\caption{{\bf Figure 4.}  The composition profile, superadiabatic
temperature gradient (radiative and actual), and mixing rate for the
$30\msun$ model for the three mixing schemes at the time when hydrogen
has burned to $X=0.2$.  Squares indicate mass bins identified as
semiconvective, for which mixing times are of order $10^6\,\rm yr$.
The stars indicate convective regions, where mixing is virtually
instantaneous.}
\endfigure

\begintable*{1}
\caption{{\bf Table 1.} $30\rm M_\odot$ Evolution.}
\catcode`\@=\active \def@{\hphantom{0}}


\halign{
$#$\hfil\qquad&\quad\hfil#\hfil\quad&\quad\hfil#\hfil\quad&
\quad\hfil#\hfil\quad\cr
\multispan4\hfil$\ell=0.5\hp$\hfil \cr
                         &  A   &  B   &  C  \cr
\tau_{\rm H}/10^6\rm yr & 5.60 & 5.27 & 5.18 \cr
\tau_{\rm He}/10^6\rm yr& 0.70 & 0.92 & 0.65 \cr
M_{\rm H}/\rm M_\odot   & 9.3@ & 9.6@ & 9.7@ \cr
M_{\rm He}/\rm M_\odot  & 7.2@ & 6.2@ & 6.6@ \cr
\noalign{\bigskip}
\multispan4\hfil$\ell=1.5\hp$\hfil \cr
                               &  A   &  B   &  C   \cr
\tau_{\rm H}/10^6\rm yr & 5.60 & 5.19 & 5.32 \cr
\tau_{\rm He}/10^6\rm yr& 0.70 & 0.89 & 0.69 \cr
M_{\rm H}/\rm M_\odot   & 9.3@ & 9.6@ & 9.7@ \cr
M_{\rm He}/\rm M_\odot  & 7.2@ & 5.8@ & 6.3@ \cr
\noalign{\bigskip}
\multispan4\hfil$\ell=5.0\hp$\hfil \cr
                               &  A   &  B   &  C   \cr
\tau_{\rm H}/10^6\rm yr & 5.60 & 5.31 & 5.33 \cr
\tau_{\rm He}/10^6\rm yr& 0.70 & 0.91 & 0.75 \cr
M_{\rm H}/\rm M_\odot   & 9.3@ & 9.6@ & 9.7@ \cr
M_{\rm He}/\rm M_\odot  & 7.2@ & 5.8@ & 6.1@ \cr
}

\endtable

We evolve a $30\msun$ star using three variations of the stellar
evolution code.  In Case A, we use only the composition mixing of the
original code.  We do, however, use the new MLT routine to determine
the temperature gradient.  This evolution is essentially identical to
the evolution of the unaltered code, and serves as our reference
model.  Case B uses the new mixing routine, where convective and
semiconvective zones are mixed at the rates computed in the new MLT
routine.  Mixing in both convective and semiconvective regions is
significantly faster than the mixing in the original code.  In Case C,
we ignore the composition gradient in the MLT routine.  This is
equivalent to using the Schwarzschild criterion for convection.  There
are no semiconvective bins, and those bins that would have been
semiconvective instead are mixed at the much faster convective rate.
The H- and He-burning times, $\tauh$ and $\tauhe$ and core masses,
$\mh$ and $\mhe$, at the times of core H exhaustion and core He
exhaustion are presented in Table 1 for mixing lengths
$\ell=0.5\hp,\,1.5\hp,\,5.0\hp$.

Figure 3 shows the evolution of the convective structure of the
$30\msun$ star.  Figure 4 shows the composition profile, temperature
gradient, and mixing times at the instant when the central composition
$X=0.2$.  In Case A, during core H-burning, there is a thin
semiconvective shell bordering on the convective core.  Slightly
separated from the convective core is a growing SCZ.  (This gap is
probably a numerical artifact, since a higher resolution calculation
does not show one.)  When only the original diffusion scheme operates,
mixing is slow enough in the SCZ that convective neutrality is not
reached, although Figure 4 shows that the departure from neutrality is
small.  As a consequence, the SCZ grows as a connected region
throughout the phase of core H-burning.  Occasionally the
semiconvective region, which usually is slightly detached from the
convective core, makes contact with the core, but the slow mixing does
not mix significant unburnt material into the interior.  This case
resembles most closely the evolution in Langer et al. (1985) where
semiconvection is suppressed ($\al=0.01$).

In Case B where the composition is mixed at the rate determined by the
extended MLT, the composition of the SCZ is able to make the
readjustments necessary to become neutrally stable and turn off the
semiconvection.  In Figure 3 the effect is to break up the SCZ into
smaller zones of semiconvection that form and disappear.  Not
surprisingly, this evolution is most similar to Langer et al.'s (1985)
evolution with $\al=1$, since we do not scale the semiconvective
mixing rate by any coefficient.  As seen in Table 1, semiconvection is
effective enough to mix some material into the core and increase the
mass of the core at H exhaustion.  A consequence of the larger core
mass is a shorter H-burning time, $\tauh$, and a slightly more
luminous main sequence phase.  The He-burning core mass, $\mhe$, is
smaller and $\tauhe$ longer for reasons discussed below.
Interestingly, Langer et al. (1985) found the opposite result, that an
enlarged H core leads to a longer $\tauh$ and a shorter $\tauhe$.
Although we are uncertain why this is, one possibility is that in
Langer et al.'s case, mass is accreted onto the core mainly near the
end of the main sequence, so that extra fuel can only extend the main
sequence lifetime, whereas in our case, mass is accreted sooner, so
that the evolution more closely resembles the evolution of a more
massive star.

Case C has the fastest mixing in the SCZ, with mixing at the
convective rate.  In the SCZ, convective bins tend to propagate inward
because mixing increases the H abundance at smaller radii, thereby
inducing instability.  The H abundance decreases at large radii,
causing stability.  The effect is akin to a rain of H enriched
material onto the convective core, and is like the episodic accretion
onto the He core of a horizontal branch star from the surrounding SCZ
(Sweigart \& Renzini 1979).  As seen in Figure 4, the composition
profile in the SCZ is not so smooth as in Case A or B.  Lamb, Iben, \&
Howard (1976), however, were able to maintain convective neutrality
and a smooth profile in the SCZ by using sufficiently fine gridding.
Case C models have the largest core mass $\mh$ at the end of
H-burning.  The core H-burning time, $\tauh$, is again shorter than in
Case A.

Only Case A evolution exhibits a blue loop in the H-R diagram, and
this is revealed in Figure 3 by the formation, retreat, and
reestablishment of a fully convective envelope during core He-burning.
In Cases B and C, core He-burning occurs mainly while the star is a
blue supergiant.  The envelope does not become fully convective until
near the end of core He-burning.  The way this evolution proceeds has
mainly to do with the structure of the intermediate convection zone
(ICZ) that forms immediately following core H exhaustion (Lauterborn,
Refsdal, \& Weigert 1971).  Following the end of core H-burning, an
ICZ is established outside the formerly convective core.  In Case A,
this zone is relatively small in extent (cf. Fig. 3), and there is
also an outer composition plateau (resulting from the convective
``finger'' at time $5.6-5.8\times 10^6\rm yr$ and mass $24-14\msun$).
In Cases B and C, however, because of the more effective mixing by
both semiconvection and convection, the ICZ and outer convective
plateau that were disconnected in Case A become connected, making
these ICZs as large as the formerly semiconvective zone.

In Cases A, B, and C, H shell burning occurs initially at the base of
the H discontinuity created by the ICZ, and since the shell is fed
efficiently by the overlying convection, the star stabilizes as a blue
supergiant, and a fully convective envelope does not form immediately.
\footnote{*}{As a general rule, an H-burning shell fed by convection
is stable as a blue supergiant, and one fed by a stable zone is a red
supergiant (Simpson 1971; Lauterborn, Refsdal, \& Weigert 1971).  A
shell fed by convection is thinner and generates more luminosity than
one at the base of a stable zone because of the substantial $\Delta X$
at the core/shell boundary.  The H-depleted cores of these blue
supergiants are smaller, therefore.  Shells fed by a stable zone with
a composition gradient must be thicker.  For these shells to generate
a comparable luminosity, they must be hotter, and, therefore, the core
has a smaller radius.  Then, by the ``mirror principle,'' the envelope
expands to red giant dimensions.}  In Case A, the small ICZ soon
disappears and the H shell burning becomes less efficient since the
shell is convectively stable.  The H shell luminosity decreases, a
fully convective envelope forms above the formerly convective ICZ, and
the star becomes a red supergiant (cf. Stothers \& Chin 1975).  When
the H shell burns through the former ICZ to the base of the convective
envelope, H shell burning is again fed by convection and the star
returns to the blue.  A secondary ICZ forms, but eventually the shell
luminosity drops slightly as the shell gets to sufficiently large
radius and a stable region forms above the shell.  This triggers a
return to the red, and soon after the H shell luminosity falls to
zero.

In Cases B and C, the ICZ is large enough that the H-burning shell
cannot burn through the composition plateau in the lifetime of the
star.  The shell is fed by convection and the star remains a blue
supergiant, gradually getting redder, for most of core He-burning, in
agreement with the $30\msun$ evolution of Simpson (1971), Stothers \&
Chin (1976), and Langer et al. (1985).  As in Case A, when the H shell
is at sufficiently large radius, a drop in the H shell luminosity
causes the shell to be fed by a stable zone, triggering the formation
of a fully convective envelope.  Because H shell burning initially
occurs at the base of an ICZ in Cases B and C, the H-burning shell is
thinner and the growing H-depleted core is smaller, making the
He-burning core mass, $\mhe$, smaller and $\tauhe$ longer.

\beginfigure{5}\vskip20.6cm
\caption{{\bf Figure 5.}  The convective structure of the $15\msun$
model for the three schemes of convective mixing.  Except for the thin
semiconvective shell at the edge of the H- and He-burning cores, the
only semiconvective region occurs in Cases A and B during the brief
period between core H-burning and core He-burning outside the formerly
convective core.  In Cases B and C, mixing during this time causes the
formation of an ICZ, the extent of which determines when the star
becomes a red supergiant.  Dotted contours outline convective regions,
and solid contours indicate semiconvection.}
\endfigure

\beginfigure{6}\vskip20cm
\caption{{\bf Figure 6.}  The composition profile, superadiabatic
temperature gradient (radiative and actual), and mixing rate for the
$15\msun$ model for the three mixing schemes at the time where core
He-burning is just beginning ($Y=0.97$).  Squares indicate mass bins
identified as semiconvective, for which mixing times are of order
$10^4\,\rm yr$.  The stars indicate convective regions, where mixing
is virtually instantaneous.  A composition plateau where an ICZ fed
the H-burning shell is apparent for Cases B and C.  Note that the
H-depleted core is largest in Case A and smaller in Cases B and C.
Consequently, Cases B and C have smaller He-burning cores, $\mhe$, and
longer $\tauhe$.}
\endfigure

\begintable*{1}
\caption{{\bf Table 2.} $15\rm M_\odot$ Evolution.}
\catcode`\@=\active \def@{\hphantom{0}}


\halign{
$#$\hfil\qquad&\quad\hfil#\hfil\quad&\quad\hfil#\hfil\quad&
\quad\hfil#\hfil\quad\cr
\multispan4\hfil$\ell=0.5\hp$\hfil \cr
                         &  A   &  B   &  C  \cr
\tau_{\rm H}/10^7\rm yr & 1.112 & 1.087 & 1.072 \cr
\tau_{\rm He}/10^7\rm yr& 0.224 & 0.259 & 0.221 \cr
M_{\rm H}/\rm M_\odot   & 2.80@ & 2.80@ & 2.80@ \cr
M_{\rm He}/\rm M_\odot  & 2.35@ & 2.20@ & 1.95@ \cr
\noalign{\bigskip}
\multispan4\hfil$\ell=1.5\hp$\hfil \cr
                               &  A   &  B   &  C   \cr
\tau_{\rm H}/10^7\rm yr & 1.112 & 1.087 & 1.072 \cr
\tau_{\rm He}/10^7\rm yr& 0.224 & 0.262 & 0.239 \cr
M_{\rm H}/\rm M_\odot   & 2.80@ & 2.80@ & 2.80@ \cr
M_{\rm He}/\rm M_\odot  & 2.35@ & 2.25@ & 1.90@ \cr
\noalign{\bigskip}
\multispan4\hfil$\ell=5.0\hp$\hfil \cr
                               &  A   &  B   &  C   \cr
\tau_{\rm H}/10^7\rm yr & 1.112 & 1.085 & 1.070 \cr
\tau_{\rm He}/10^7\rm yr& 0.224 & 0.264 & 0.249 \cr
M_{\rm H}/\rm M_\odot   & 2.80@ & 2.80@ & 2.80@ \cr
M_{\rm He}/\rm M_\odot  & 2.35@ & 2.25@ & 1.90@ \cr
}

\endtable

We consider how these results depend on the mixing length $\ell$.
Case A is effectively independent of the mixing length.  The mixing
length, $\ell$, only bears on the temperature gradient of the
convective core, which is so close to adiabatic for any reasonable
value that the interior evolution is the same.  According to equation
(14), the semiconvective diffusion rate does not depend on $\ell$, so
that Case B main sequence evolution is independent of $\ell$ within
numerical uncertainty.  Following core H depletion, the ICZ forms more
quickly for large $\ell$, however, making the H-depleted core and
$\mhe$ somewhat smaller for larger values of $\ell$.  This effect is
more clearly illustrated in Case C.  The main sequence evolution is
hardly affected by the choice of $\ell$, but the faster formation of
the ICZ for larger $\ell$ reduces $\mhe$ and increases $\tauhe$.

\subsection{$15\msun$ Evolution}

We evolve a $15\msun$ star for the same three convection schemes
described above.  The evolution of the convective structure is shown
in Figure 5, and the composition and temperature profiles at the onset
of core He-burning ($Y=0.97$) are shown in Figure 6.  The times
$\tauh$ and $\tauhe$ and corresponding core masses, $\mh$ and $\mhe$,
are listed in Table 2.

As seen in Figure 5, there is no semiconvective region during core
H-burning, except for the thin shell surrounding the core.  The only
semiconvection that is important for the evolution occurs during the
brief interval between core H-burning and core He-burning, outside the
formerly convective core.  Although evolution through this phase
occurs on the fast thermal time scale, typical semiconvective time
scales of order $10^4\,\rm yr$ are fast enough to effect the internal
structure of the star.  In Case A, where only the original mixing is
used, semiconvection does not cause any significant mixing outside the
formly convective core.  Because convection effectively does not feed
the H-burning shell, the shell must be relatively thick to generate
the required luminosity.  In Cases B and C, an ICZ creates a
composition plateau.  The ICZ feeds the H-burning shell, which is
thinner than in Case A.  Thus, in Figure 6 the H-depleted core has a
smaller mass in Case B, so that the He-burning core mass is smaller.
The H-depleted core and mass $\mhe$ are even smaller in Case C.  The
He-burning time $\tauhe$ is shortest in Case A.

The $15\msun$ model becomes a red giant when the ICZ disappears.  For
all mixing schemes, the star remains a red giant throughout core
He-burning.  This is in contrast to most other results, where the star
either starts evolution as a blue supergiant (Stothers \& Chin 1976;
Simpson 1971; Lamb et al. 1976) or makes a blue loop (Stothers \& Chin
1975; Langer et al. 1985), returning to the red only after central
convection has ceased.

As with the $30\msun$ model, Case A is independent of the mixing
length, and Case B nearly so.  The value of the mixing length is most
important for Case C evolution.  In this case a larger $\ell$ makes
the ICZ more efficient at delivering fuel to the H-burning shell,
making the shell thinner.  The subsequent He core mass, $\mhe$, is
slightly smaller, and $\tauhe$ is slightly longer.

\section{Summary and Discussion}

We have derived the extended local MLT for fluids with composition
gradients.  Our method has been to solve the local moment equations of
GNA.  As is well-known from linear stability analyses, fluids with
both temperature and composition gradients can experience
double-diffusive instabilities, where competing diffusion rates
determine the nature of the instability.  The salt finger and
semiconvective instabilities grow much more slowly than the dynamical
convective instability, at a rate proportional to the radiative
diffusion.  Our extended MLT provides the characteristic turbulent
velocities at which the linear instabilities saturate.  This nonlinear
theory applies to all stability regimes--convective, semiconvective,
salt finger, and stable.  Standard MLT for a fluid of homogeneous
composition is a limiting case.

We have demonstrated that although the precise stability criteria
require solving equations (6)--(8), the familiar Schwarzschild and
Ledoux criteria are adequate approximations for stellar interiors.
Likewise, although calculation of the precise mixing rate requires
solving equation (9), the approximate solutions given by equations
(12) for convection, (13) for salt finger mixing, and (14) for
semiconvective mixing are good in most circumstances.  The rate of
semiconvective diffusion does not depend on the value of the mixing
length.

We have written a subroutine that solves the extended MLT for the
self-consistent temperature gradient and mixing rate, and used it in
Eggleton's stellar evolution code.  It has also been necessary to
supply a subroutine that permits faster mixing than the unmodified
code allows.  We have evolved $30\msun$ and $15\msun$ stars from the
start of core H-burning to the end of of core He-burning.  The results
presented here mainly confirm results found previously by others (cf.
the review of massive star evolution by Chiosi \& Maeder 1986).  The
$30\msun$ star has a semiconvective region outside the H-burning
convective core.  The $10^6\,\rm yr$ mixing time scale is fast enough
that semiconvection maintains the composition profile at convective
neutrality, so that semiconvection is continually initiated and
quenched during main sequence evolution.  This broken-up structure of
the SCZ most closely resembles the evolution of Langer et al. (1985)
with $\al=1$.  Of greater importance for subsequent evolution is the
development of an ICZ during the brief period after core H-burning
stops and before core He-burning begins.  Semiconvection allows the
formation of a substantial ICZ, causing He-burning to begin while the
star is a blue supergiant.  The star does not become a red supergiant
until near the end of core He-burning.

Although the $15\msun$ star does not have an extended semiconvective
zone during main sequence evolution, it does have one during the brief
time between core H- and core He-burning.  The treatment of mixing at
this time determines the evolution of the ICZ, which in turn
determines evolution during later stages.  Core He-burning occurs
mainly in the red and makes no blue loops, in contrast to most other
authors, who find that core He-burning either occurs mostly in the
blue, or makes a blue loop if it starts in the red.  It has not been
our intention to perform state-of-the-art evolutionary calculations
for stars of these masses, and we are not making claims about the
formation of blue loops.  Our aim has been to demonstrate the
feasibility of using the extended MLT in stellar evolution
calculations and to show that semiconvective mixing time scales are in
the interesting range where mixing is important on evolutionary time
scales, but not so fast that instantaneous readjustment of composition
is appropriate.

Horizontal branch (HB) stars also develop semiconvective zones
(Paczy\'nski 1970; Schwarzschild 1970).  Increasing C/O abundance in
the He-burning core increases the opacity.  Thus, during the early
stages of evolution, mixing at the core boundary causes instability
just outside the core, and the core grows in an overshooting phase
(Castellani, Giannone, \& Renzini 1971a).  Eventually a time comes
when an unstable, high opacity region forms around the convective core
(in this case due to C/O enrichment), that is, beyond the radius of
Schwarzschild stability, much as the $30\msun$ star does during main
sequence evolution.  This region is thought to experience a
readjustment of composition, where mixing with the He-rich material of
the envelope restores convective neutrality.  This region of partial
mixing is the semiconvective zone of HB stars (Castellani, Giannone,
\& Renzini 1971b).  The prescription for maintaining instantaneous
neutrality in this region is the ``canonical semiconvective scheme''
commonly used in horizontal branch evolution.  An algorithm has been
described by Robertson \& Faulkner (1972), and a recent discussion can
be found in Dorman \& Rood (1993).  The ratio of the number of AGB to
HB stars depends strongly on the rate of mixing in the SCZ (Renzini \&
Fusi-Pecci 1988), and evolution with the canonical scheme is
consistent with observed ratio (Buzzoni et al. 1983).  It would be
interesting to verify whether or not mixing using the extended MLT is
consistent with the observations, whether the canonical scheme is
appropriate, and whether there are any consequences to the finite rate
of mixing in the SCZ.

An issue we have not addressed so far is the possibility of nonlocal
semiconvection.  That is, do fast convective velocities penetrate far
into adjacent semiconvective regions (convective overshooting)?  Do
slow semiconvective velocities slowly mix into adjacent stable regions
(semiconvective overshooting, cf. Aur\'e 1971)?  Answers to these
questions would require that we develop the nonlocal moment equations
for fluids with composition gradients.  GNA developed the nonlocal
equations for fluids of homogeneous composition, and it would be
straight-forward to extend those results to this more complicated
problem.  The nonlocal moment equations are likely to be quite
difficult to solve, however.  This problem requires solving a large
number of coupled differential equations, rather than the algebraic
equations of the local theory.  Experience has shown that solutions to
the simpler problem of nonlocal convection in a homogeneous fluid are
difficult to obtain (Grossman 1996).  In principle, one could also
simulate nonlocal mixing in a fluid with a composition gradient using
the GSPH technique of Grossman \& Narayan (1993).  The current
consensus is that composition gradients are effective barriers to
rapid mixing (cf. Shibahashi \& Osaki 1976), and we think this is
likely to be proved correct when more detailed results are available.
We think it possible, however, that slow semiconvective mixing may
extend significantly into stable zones.

Finally, we emphasize again a central shortcoming of all the preceding
discussion, that the theory of semiconvective mixing proposed here has
all the same shortcomings as standard mixing length theory.  In
particular, the effective viscosity of the turbulent cascade of a
fluid at high Reynolds number is approximated by an eddy-damping rate,
a rate that accounts for turbulent dissipation at only the largest
scale.  This rate depends on a single, unknown parameter--the mixing
length.  Stevenson (1979) has proposed a mechanism for semiconvection
where linear growth develops on the scale of maximum instability,
which is much smaller than a pressure scale height, and energy
cascades into even smaller scales.  This raises the question of
whether an appropriate mixing length for semiconvection is of order a
pressure scale height or is much smaller.  We have shown that if
motions are dominated by the scale of maximum instability or larger,
the rate of semiconvective diffusion is independent of $\ell$ and
bigger than if motions are dominated by the smaller scales.  Thus, in
semiconvective regions, one can continue to use the same, large mixing
length as in convective regions; it is not necessary to adopt a
smaller value.

A more accurate treatment of semiconvection requires solving the
multidimensional hydrodynamic equations.  The first 2-dimensional
simulation of semiconvection has been presented by Merryfield (1995).
Unfortunately, the vertical depth of the simulation is comparable to
the size of maximum growth, and is much smaller than the size of
semiconvective regions in stars, thus giving little information about
composition transport over distances comparable to a pressure scale
height.  Even if one could simulate semiconvection in a fluid of
appropriate depth, the simulation would represent only one instant in
the evolution of a star; one cannot evolve a star using
multidimensional hydrodynamics.  Consequently, highly-simplified
prescriptions for mixing, such as MLT, will continue to be useful for
a long time yet.

\section*{Acknowledgments} It is with pleasure that we thank Ramesh
Narayan for his participation in the development of the moment theory
of convection, which is central to the ideas and methods that were
used here.  S.A.G. is grateful to Brian Chaboyer for discussions of
the importance of semiconvection for stellar evolution.  Without this
early motivation, the evolutionary calculations presented here might
not have been attempted.  S.A.G. also thanks Susan Lamb for
discussions of past work and for pointing out certain references.
S.A.G. is supported under NASA grant NAGW-2935.  R.E.T. has been
supported in part by NSF under grant AST94-15423.
\bigskip

\def\apj{ApJ}

\def\apjs{ApJ Supp}

\def\mnras{MNRAS}
\def\qjras{QJRAS}
\def\aa{A\&A}

\def\annrev{ARA\&A}

\def\pasj{PASJ}

\def\rhang{\noindent\hangindent\parindent\hangafter1}

\def\rj#1#2#3#4{\rhang#1, #2, #3, #4\par}
\def\rb#1#2#3{\rhang#1, #2, \rm\ #3\par}

\beginrefs

\rj{Arnett, D. 1991}{\apj}{383}{295}
\rj{Aur\'e, J.-L. 1971}{\aa}{11}{345}
\rj{Baines, P.G., \& Gill, A.E. 1969}{J. Fluid Mech.}{37}{289}
\rj{B\"ohm-Vitense, E. 1958}{Z.~Astrophys.}{46}{108}
\rj{Buzzoni, A., Fusi Pecci, F., Buonanno, R., \& Corsi, C.E. 1983}
{\aa}{128}{94}
\rj{Castellani, V., Giannone, P., \& Renzini, A. 1971a}{Ap \& SS}{10}{340}
\rj{Castellani, V., Giannone, P., \& Renzini, A. 1971b}{Ap \& SS}{10}{355}
\rj{Chiosi, C., \& Maeder, A. 1986}{\annrev}{24}{329}
\rj{Dorman, B., \& Rood, R.T. 1993}{\apj}{409}{387}
\rj{Eggleton, P.P. 1971}{\mnras}{151}{351}
\rj{Eggleton, P.P. 1972}{\mnras}{156}{361}
\rj{Eggleton, P.P. 1973}{\mnras}{163}{279}
\rj{Eggleton, P.P. 1983}{\mnras}{204}{449}
\rj{Gabriel, M. 1969}{\aa}{1}{321}
\rj{Gabriel, M. 1970}{\aa}{6}{124}
\rj{Gabriel, M., \& Noels, A. 1976}{\aa}{53}{149}
\rj{Gough, D.O., \& Toomre, J. 1982}{J. Fluid Mech.}{125}{75}
\rj{Grossman, S.A., Narayan, R., \& Arnett, D. 1993}{\apj}{407}{284 (GNA)} 
\rj{Grossman, S.A., \& Narayan, R., 1993}{\apjs}{89}{361} 
\rj{Grossman, S.A. 1996}{\mnras}{279}{305} 
\rj{Kato, S. 1966}{PASJ}{18}{374}
\rb{Kippenhahn, R., \& Weigert, A. 1990}{Stellar Structure and Evolution}
{New York: Springer}
\rj{Kippenhahn, R., Ruschenplatt, G., \& Thomas, H.-C. 1980}{\aa}{91}{181}
\rj{Lamb, S.A., Iben, I., Jr., Howard, W.M. 1976}{\apj}{207}{209}
\rj{Langer, N. 1991}{\aa}{252}{669}
\rj{Langer, N., Sugimoto, D. \& Fricke, K.J. 1983}{\aa}{126}{207}
\rj{Langer, N., El Eid, M.F. \& Fricke, K.J. 1985}{\aa}{145}{179}
\rj{Langer, N., El Eid, M.F. \& Baraffe, I. 1989}{\aa}{224}{L17}
\rj{Lauterborn, D., Refsdal, S., \& Weigert, A. 1971}{\aa}{10}{97}
\rj{Ledoux, P. 1947}{\apj}{105}{305}
\rj{Merryfield, W.J. 1995}{\apj}{444}{318}
\rj{Nakakita, T., \& Umezu, M. 1994}{\mnras}{271}{57}
\rj{Paczy\'nski, B. 1970}{Acta Astron}{20}{195}
\rj{Proctor, M.R.E. 1981}{J. Fluid Mech.}{105}{507}
\rj{Renzini, A., \& Fusi Pecci, F. 1988}{\annrev}{26}{199}
\rj{Robertson, J.W., \& Faulkner, D.J. 1972}{\apj}{171}{309}
\rj{Sakashita, S. \& Hayashi, C. 1959}{Prog. Theo. Phys}{22}{830}
\rj{Schwarzschild, M., \& H\"arm, R. 1958}{\apj}{128}{348}
\rj{Schwarzschild, M. 1970}{\qjras}{11}{12}
\rj{Shibahashi, H., \& Osaki, Y. 1976}{\pasj}{28}{199}
\rj{Simpson, E.E. 1971}{\apj}{165}{295}
\rj{Spiegel, E.A. 1969}{Comm. Astrophys. and Space Phys.}{1}{57}
\rj{Spiegel, E.A. 1972}{\annrev}{10}{261}
\rj{Spruit, H.C. 1992}{\aa}{253}{131}
\rj{Stevenson, D.J. 1977}{Proc.~Astron.~Soc.~Australia}{3}{165}
\rj{Stevenson, D.J. 1979}{\mnras}{187}{129}
\rj{Stothers, R.B., \& Chin, C.-w. 1975}{\apj}{198}{407}
\rj{Stothers, R.B., \& Chin, C.-w. 1976}{\apj}{204}{472}
\rj{Stothers, R.B., \& Chin, C.-w. 1992a}{\apj}{390}{136}
\rj{Stothers, R.B., \& Chin, C.-w. 1992b}{\apj}{390}{L33}
\rj{Stothers, R.B., \& Chin, C.-w. 1994}{\apj}{431}{797}
\rj{Sweigart, A.V., \& Renzini, A. 1979}{\aa}{71}{66}
\rj{Ulrich, R.K. 1972}{\apj}{172}{165}
\rj{Umezu, M., \& Nakakita, T. 1988}{Ap \& SS}{150}{115}
\rj{Umezu, M. 1989}{Ap \& SS}{162}{13}
\rj{Xiong, D. 1981}{Scientia Sinica}{24}{1406}

\endrefs
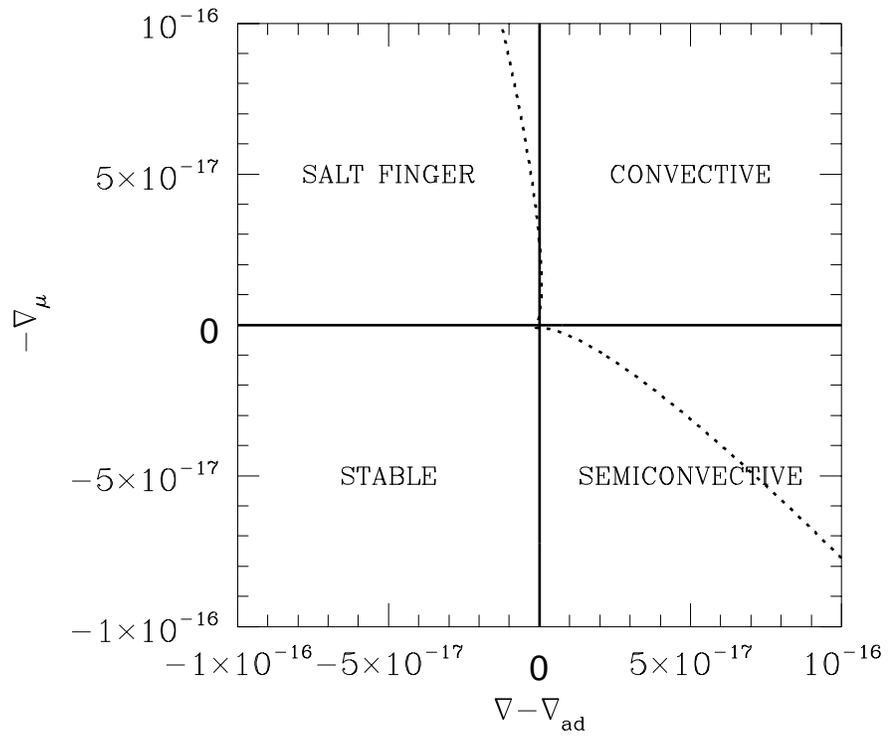
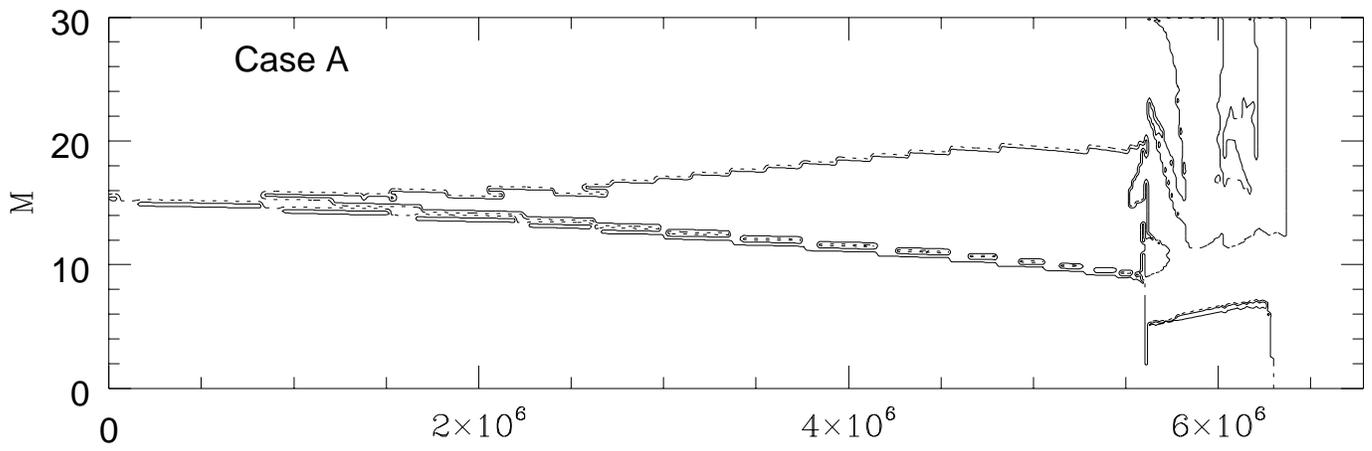
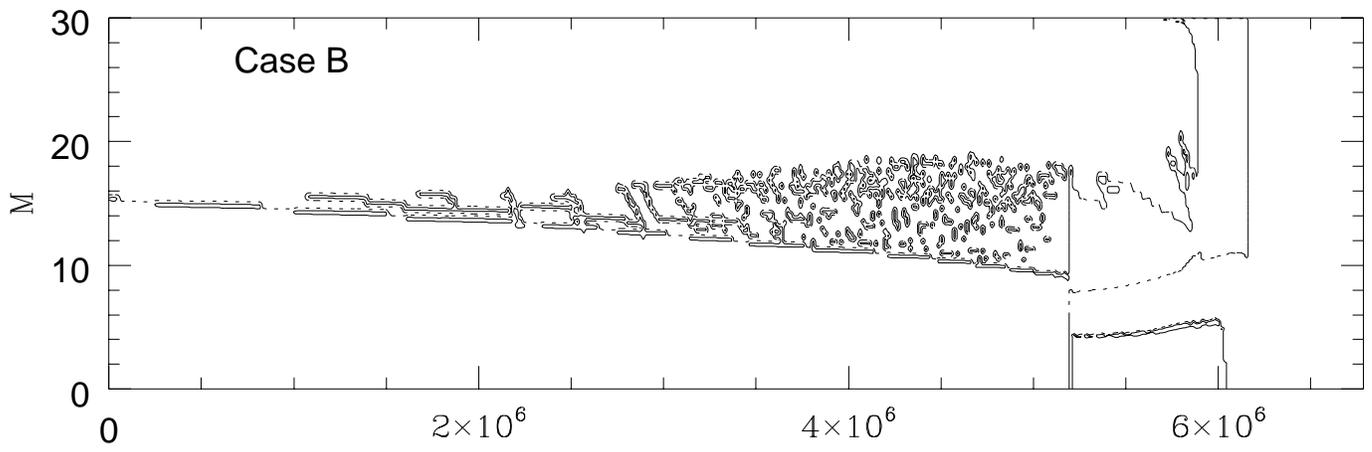
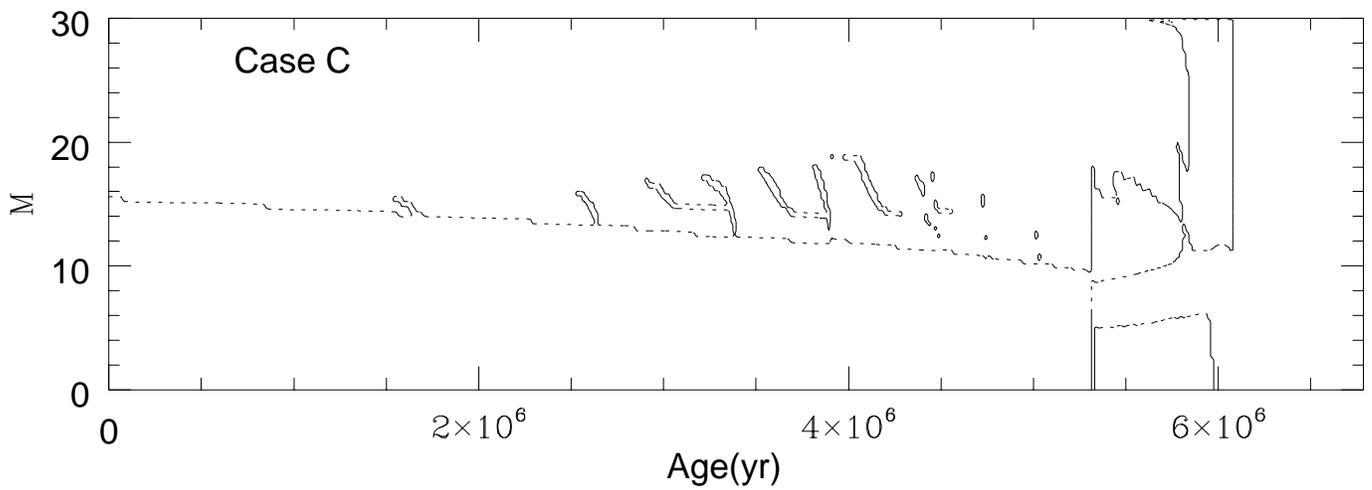
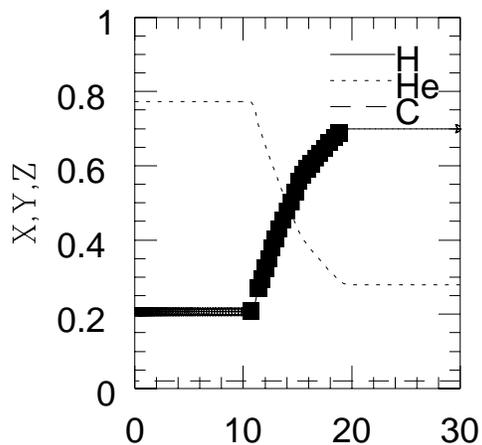
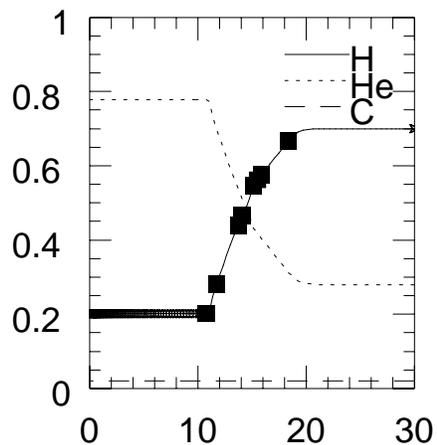
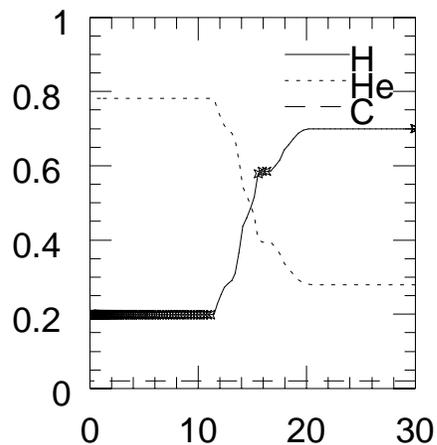
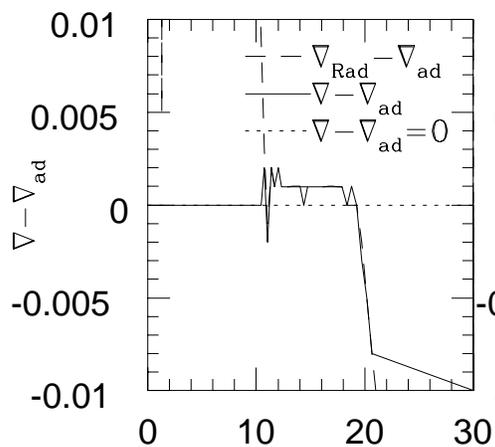
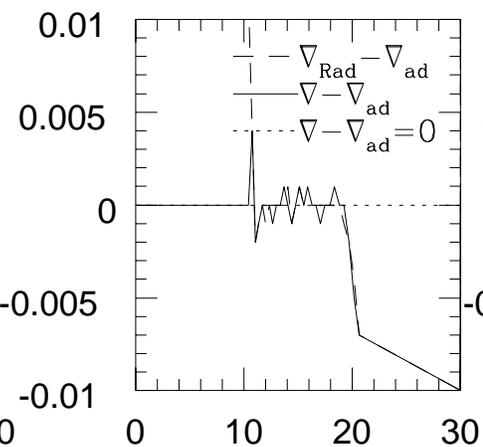
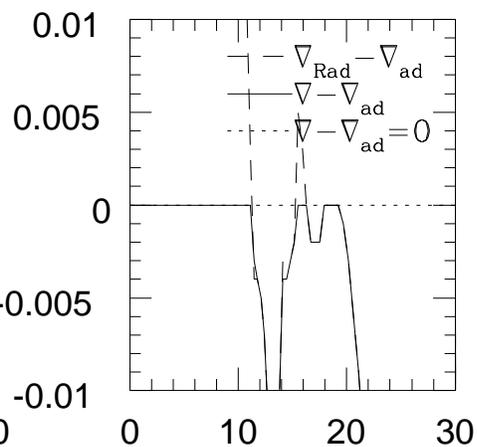
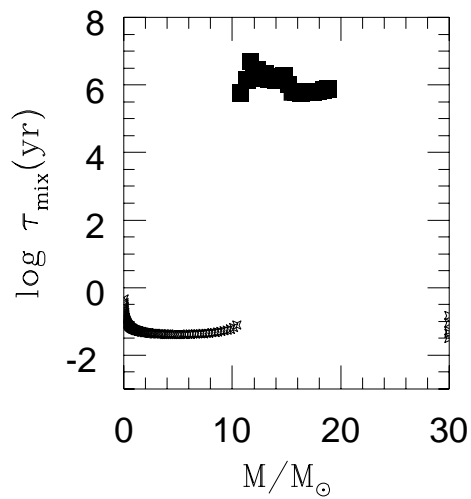
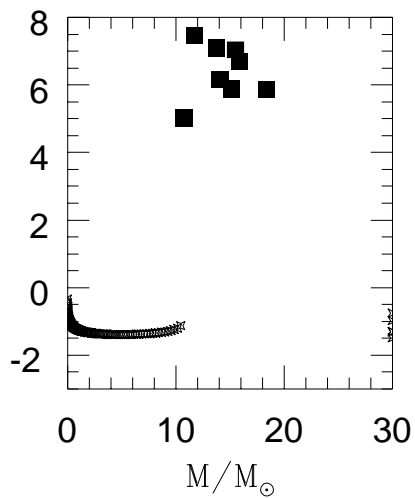
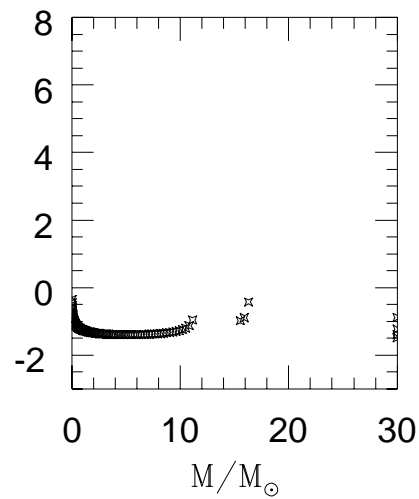
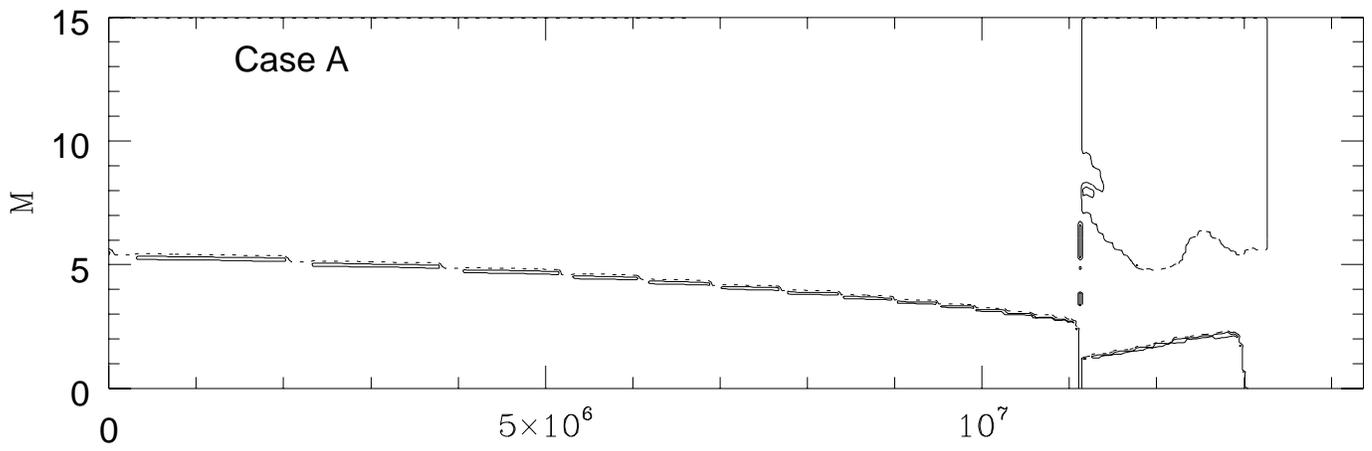
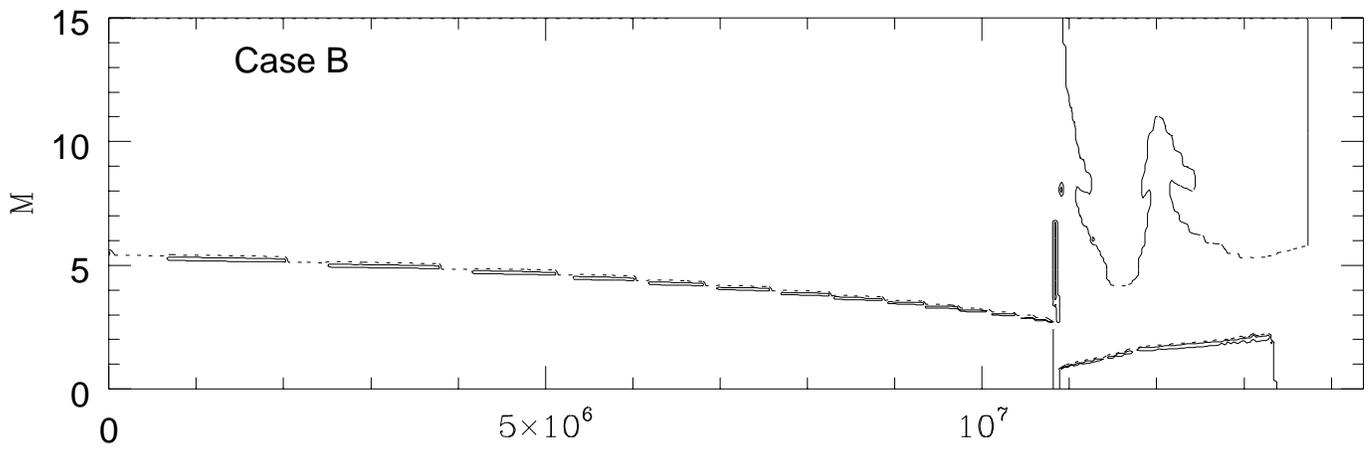
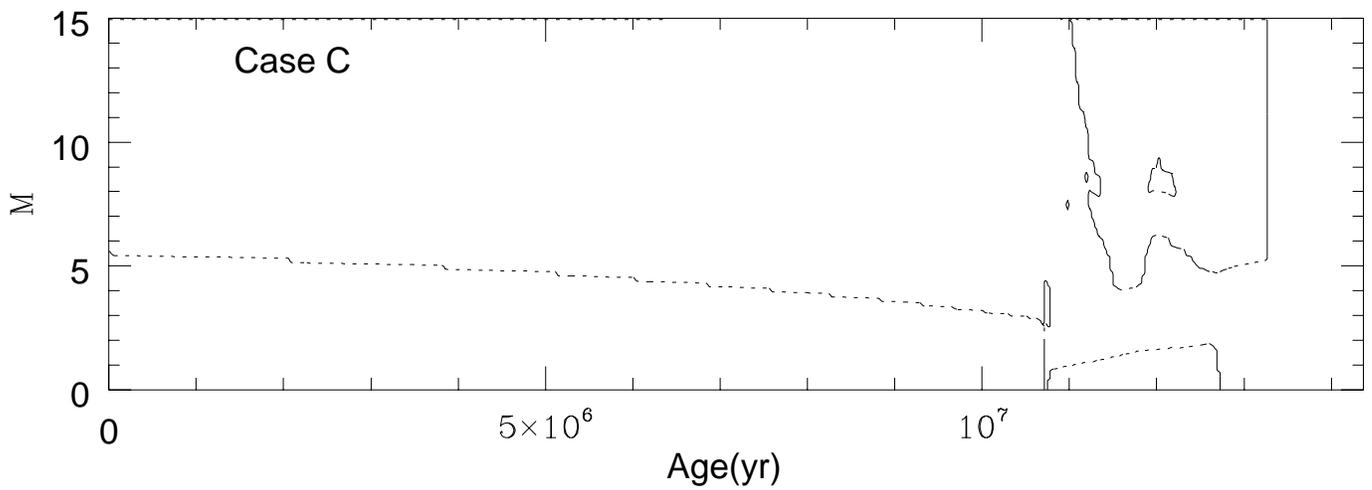
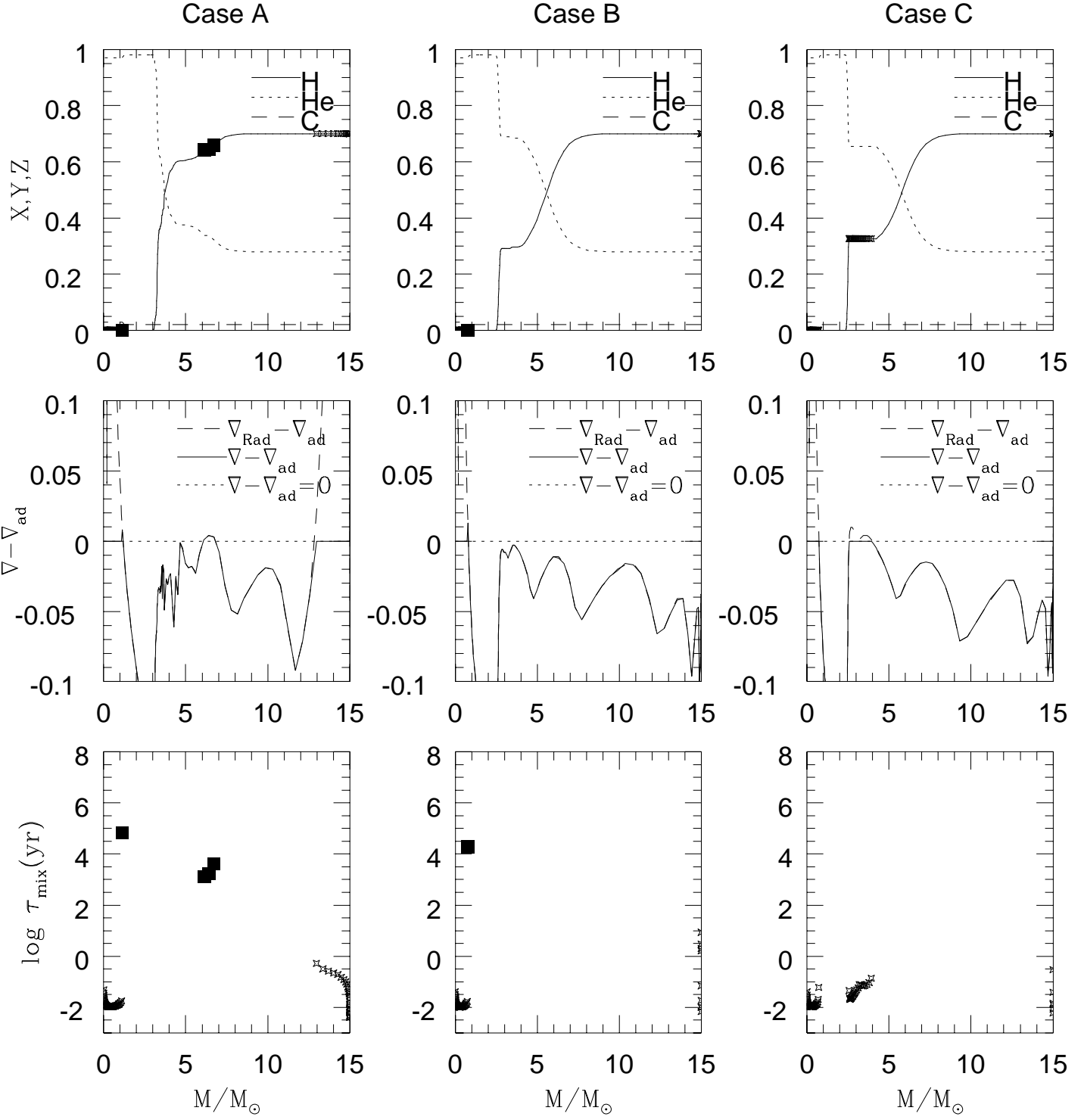
\bye